\documentclass[onecolumn,showkeys,aps,prd]{revtex4}

\usepackage{amsfonts}
\usepackage{latexsym}
\usepackage{multirow}
\usepackage{tikz}
\usepackage{amsmath,amssymb,amsthm}
\usepackage{mathtools}
\usepackage{lineno}
\usepackage{bbm}
\usepackage{float}      

\usetikzlibrary{arrows.meta}
\usetikzlibrary{calc}

\usepackage{stackengine,graphicx}

\usepackage{xcolor}
\usepackage[normalem]{ulem}

\renewcommand{\dj }{\mbox{\kern0.6ex\raise0.8ex\hbox{-}\kern-1.4ex d}}
\newcommand{\Dj }{\mbox{\raise0.3ex\hbox{-}\kern-0.4em D}}

\newcommand{\ds}{\displaystyle}

\newcommand{\sspan}{\mathop{\rm span}\nolimits }
\newcommand{\diag}{\mathop{\rm diag}\nolimits }

\newcommand{\realni}{\ensuremath{\mathbb{R}}}

\newcommand{\cN}{{\cal N}}

\renewcommand{\leq}{\leqslant}
\renewcommand{\geq}{\geqslant}

\begin{document}

\title{Minkowski metric signature as a statistical\\ prediction of piecewise-flat quantum gravity}

\author{Mihailo \Dj or\dj evi\'c}
 \email{mdjordjevic@ipb.ac.rs}
\affiliation{Institute of Physics, University of Belgrade, Pregrevica 118, 11080 Belgrade, Serbia}

\author{Igor Prlina}
 \email{prlina@ipb.ac.rs}
\affiliation{Institute of Physics, University of Belgrade, Pregrevica 118, 11080 Belgrade, Serbia}

\author{Marko Vojinovi\'c}
 \email{vmarko@ipb.ac.rs}
 \thanks{corresponding author}
\affiliation{Institute of Physics, University of Belgrade, Pregrevica 118, 11080 Belgrade, Serbia}



\begin{abstract}
We study an \emph{ab initio} calculation of the signature of the spacetime metric, within the framework of piecewise-flat models of quantum gravity. Applying numerical techniques, we evaluate the probability that any $4$-simplex in a triangulation of a spacetime manifold has a given metric signature. For this we use a toy-model of quantum gravity based on quantum Regge calculus, where we ignore any effects due to dynamics and focus only on the kinematics of the configuration space. Our results may hold even for other QG models, as long as they behave similarly in the same approximation. We obtain that for a suitably chosen spacelike domain of edge lengths and the UV cutoff, the probability of observing Minkowski signature reaches approximately $85\,\%$. In contrast, allowing both spacelike and timelike edge lengths leads to the split signature being dominant, with probability between $72\,\%$ and $81\,\%$. The results are essentially independent of the choice of topology and triangulation of the spacetime manifold. Our findings demonstrate that the signature of the metric does not necessarily need to be postulated in a given model of quantum gravity, but can instead be predicted from it, and we estimate the relative size of the UV cutoff and the domain for edge lengths that a given model should incorporate in order to predict the Minkowski signature with high probability. There is a potential to further increase this probability by taking into account the dynamics of the theory.
\end{abstract}

\maketitle

\section{\label{SecI}Introduction}

One of the fundamental open questions of modern theoretical physics is the question of the signature of the spacetime metric. The famous ``opposite sign'' distinguishing one time direction from three spatial directions is related to fundamental questions such as what distinguishes time from space, why time is one-dimensional while space is three-dimensional, and so on. For the most part, there is no consensus on any possible theoretical explanation for this particular signature, other than it is a direct consequence of the postulates of special relativity, in particular the invariance of the speed of light.

This situation is of course less than satisfactory, and there exists substantial literature discussing the signature of the spacetime metric, in various contexts. These contexts can be roughly classified as follows:
\begin{itemize}
\item Research discussing various theoretical assumptions that require, or at least favor, the Minkowski signature over other signatures \cite{Tegmark,DamNg,BorstnikNielsen}.
\item Research discussing various experimentally verified physical phenomena that can be leveraged to establish the Minkowski signature \cite{ItinHehl,ObukhovHehl,RaetzelRiveraSchuller,Rubilar}.
\item Research discussing the mathematics of complex metric and smooth signature changes via the complex plane \cite{KontsevichSegal,Witten}.
\item Research discussing the consistency and properties of multi-time physics and the split signature \cite{Dorling,John,CraigWeinstein,Weinstein,Yndurain,BarrettGibbonsPerryPopeRuback,Bars}.
\item Research discussing the consequences of signature change in general relativity and other theories \cite{Ellis,EllisSumerukCouleHellaby,DrayManogueTucker}.
\item Research discussing signature flip in quantum cosmology \cite{Sakharov,HartleHawking,GibbonsHartle}.
\item Research discussing dynamical ways that establish the emergence of Minkowski signature from a Euclidean theory \cite{Greensite,CarliniGreensite,GirelliLiberatiSindoni,MukohyamaUzan,Mukohyama}.
\end{itemize}

While the versatility of the existing literature is substantial, there are too few papers that address the question of signature from the point of view of quantum gravity (QG), which is, in fact, arguably the most appropriate framework to discuss it. While there is also no consensus on the question of the theory of QG itself, there are many tentative models which (while still being under construction and a work in progress) may be mature enough to tackle the question of the signature of the metric. The purpose of this work is to present an \emph{ab initio} approach to the question of metric signature within the framework of the so-called \emph{piecewise-flat} models of QG \cite{MikovicVojinovicBook}.

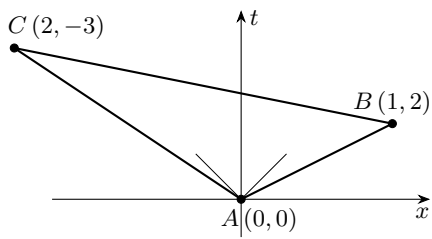
\begin{figure}[H]
\begin{center}
\begin{tikzpicture}
\draw[thick] (0,0)  -- (2,1);
\draw[thick] (0,0)  -- (-3,2);
\draw[thick] (-3,2)  -- (2,1);
\filldraw[black] (0,0) circle (1.5pt);
\filldraw[black] (2,1) circle (1.5pt);
\filldraw[black] (-3,2) circle (1.5pt);
\draw[thin,-Stealth] (-2.5,0) -- (2.5,0);
\draw[thin,-Stealth] (0,-0.5) -- (0,2.5);
\draw[very thin] (0,0) -- (0.6,0.6);
\draw[very thin] (0,0) -- (-0.6,0.6);
\node[below] at (0.24,0)  {$A\,(0,0)$};
\node[above] at (2,1) {$B\,(1,2)$};
\node[above right] at (-3.2,2) {$C\,(2,-3)$};
\node[right] at (0,2.4) {$t$};
\node[below] at (2.4,0) {$x$};
\end{tikzpicture}
\end{center}
\caption{\label{fig:MinkowskiTriangle}A triangle in Minkowski plane. Its three edges have lengths $l_{AB} = \sqrt{3} \approx 1.73$, $l_{AC} = \sqrt{5} \approx 2.23$, and $l_{BC} = \sqrt{24} \approx 4.89$. One can observe that the triangle inequalities are not satisfied, since $l_{BC} > l_{AB} + l_{AC}$. A small future-pointing light cone centered at the origin is depicted as well, for reference.}
\end{figure}

The basic intuition behind the idea of studying the signature of the metric in the context of piecewise-flat QG models stems from the following insight of elementary geometry. Take three positive numbers, $a,b,c\in(0,1]$, think of them as lengths of three line segments (i.e., edges), and ask when these three edges can form a triangle. The naive answer, typically taught in math classes, is:
\begin{quote}
\em Three edges can form a triangle if and only if their lengths satisfy the triangle inequalities,
\begin{equation} \label{eq:TriangleInequalities}
a \leq b+c\,, \qquad b \leq a+c\,, \qquad c\leq a+b\,.
\end{equation}
\end{quote}
Taken at face value, the above answer is \emph{wrong}. A more precise answer would be the following:
\begin{quote}
\em Three edges can \emph{always} form a triangle. If their lengths satisfy triangle inequalities (\ref{eq:TriangleInequalities}), the resulting triangle is embeddable in the Euclidean plane, while if their lengths fail to satisfy triangle inequalities, the resulting triangle is embeddable in the Minkowski plane.
\end{quote}
An obvious example of a triangle whose edge lengths fail to satisfy triangle inequalities is depicted in Figure \ref{fig:MinkowskiTriangle}. Its edge lengths are evaluated via the formula for Minkowski distance between the points with coordinates $(t_m,x_m)$ and~$(t_n,x_n)$:
\begin{equation}
l_{mn} = \sqrt{-(t_m-t_n)^2 + (x_m - x_n)^2}\,.
\end{equation}

The above elementary insight --- that the status of triangle inequalities provides information about the signature of the metric of the ambient space --- represents the core idea of the general method to study the signature of the spacetime metric that we develop in this paper.

In addition to satisfying or violating triangle inequalities, one can also ask a more quantitative question, namely \emph{how often} do randomly chosen lengths $a,b,c\in(0,1]$ give rise to a Euclidean or a Minkowski triangle. A simple analysis of the $3$-dimensional volume defined by inequalities (\ref{eq:TriangleInequalities}) within the unit cube in $(a,b,c)$-space shows that a randomly chosen point in this space corresponds to a Euclidean triangle with probability $1/2$, i.e., $50\,\%$, and Minkowski triangle also with probability $50\,\%$. However, this is true only if we consider the spacelike edge lengths for the Minkowski triangles. In general, however, each edge length of a Minkowski triangle may be either spacelike or timelike (note that lightlike edges have zero length and represent a degenerate case), i.e., a more general domain for the lengths is
\begin{equation} \label{eq:MixedPlaneDomain}
  a,b,c \in (0,1] \cup (0,i]\,, \qquad i \equiv \sqrt{-1}\,.
\end{equation}
Using this domain for the edge lengths, there are now \emph{three} distinct types of triangles that can appear:
\begin{itemize}
\item a Euclidean triangle, embeddable into a plane with a metric tensor $[\eta_{\mu\nu}] = \diag(1,1)$,
\item a Minkowski triangle, embeddable into a plane with metric tensor $[\eta_{\mu\nu}] = \diag(-1,1)$, and
\item a negative Euclidean triangle, embeddable into a plane with metric tensor $[\eta_{\mu\nu}] = \diag(-1,-1)$.
\end{itemize}
An example of a negative Euclidean triangle would be an equilateral triangle with edge lengths chosen as $a=b=c=i$. Given a random choice of edge lengths $a,b,c$ from the domain (\ref{eq:MixedPlaneDomain}), the probabilities of obtaining the above three types of triangles are $1/16$, $7/8$, $1/16$, or expressed in percentages, $6.25\,\%$, $87.5\,\%$ and $6.25\,\%$, respectively. Therefore, for this mixed domain for edge lengths, the Minkowski triangles are far more common than Euclidean and negative Euclidean triangles.

There is yet another question one can ask, related to the above analysis. Namely, suppose we are given two adjacent triangles, with edge lengths $(a,b,c)$ and $(a,b',c')$ respectively, sharing a common edge of length $a$. Then, if $a$ is spacelike, namely $a\in(0,1]$, and given convenient choices of $b,c,b',c'$, both triangles can be embeddable in the Euclidean plane. Also, with a different convenient choice of $b,c,b',c'$, one triangle may be embeddable into the Euclidean plane, while the other into the Minkowski plane, for example. However, if $a$ is timelike, $a\in(0,i]$, neither of the two triangles is embeddable in the Euclidean plane, regardless of the choices of $b,c,b',c'$. Both triangles are then embeddable only into the Minkowski plane (or possibly the negative Euclidean plane). This leads us to discuss the \emph{correlations} of the signatures of the metric for adjacent triangles, depending on the lengths of their common edges, in the sense that given a triangle of a certain type, the types of its neighboring triangles cannot always be chosen independently.

The main point of our work is to generalize the above discussion from a $2$-dimensional plane to a $4$-dimensional spacetime, and establish a method to evaluate the probabilities of all possible signatures of the metric tensor, given the information about edge lengths, within a framework of piecewise-flat QG models.

Namely, in the models of QG based on a smooth $4$-dimensional spacetime manifold, the signature of the metric is a topological invariant, and the only way to change it is to glue a manifold with one signature to a manifold with a different signature along a boundary where the signature is degenerate, often called a \emph{domain wall} \cite{Ellis,EllisSumerukCouleHellaby}. While these scenarios are sometimes studied in the context of quantum cosmology \cite{Sakharov,HartleHawking,GibbonsHartle}, the signature of the metric cannot be truly considered a \emph{variable} in such QG models. On the other hand, QG models based on a piecewise-flat spacetime manifold are much more flexible in this regard, since each flat cell in the piecewise-flat structure can give rise to a locally induced metric with a different signature, typically depending on the edge lengths that describe the particular cell. In this setup, the signature of the metric becomes a function of the edge lengths of an individual cell, and thus a \emph{proper variable} in the model --- given a convenient choice of edge lengths, one can discuss spacetime configurations which feature different signatures at different cells. Moreover, one can sum over all possible configurations of edge lengths, introducing a path integral that kinematically takes into account all possible signatures for all cells in the piecewise-flat manifold. Therefore, QG models based on piecewise-flat manifolds have a natural advantage over all other QG models in being able to discuss (and possibly even predict) the signature of the metric as a consequence of the kinematics and dynamics of the model.

The idea of constructing QG models on a piecewise-flat manifold goes all the way back to the classical Regge calculus \cite{ReggeCalculus}, and the Ponzano-Regge model \cite{PonzanoRegge} as the first QG model of this kind, formulated for $3$-dimensional manifolds. Various models of quantum Regge calculus have been discussed (see \cite{Hamber} for a review and further references), while within the Loop Quantum Gravity approach, various \emph{spinfoam models} have been formulated and studied \cite{RovelliVidotto2014}. The question of how to add matter fields to spinfoam models later gave rise to an additional class of generalized QG models, employing the framework of \emph{higher gauge theory} \cite{BaezHuerta}. For example, the \emph{spincube model} \cite{MikovicVojinovic2012} is based on the notion of the Poincar\'e $2$-group, while a more recent model \cite{Stipsic2025b,StipsicVojinovicQGmodel} is based on the notion of the so-called Standard Model $3$-group \cite{Radenkovic2019}.

While most of the above QG models have been constructed, for simplicity, with the imagined fixed signature of the spacetime metric, it is not really necessary to introduce this as an assumption into the model. As we demonstrate in this work, one can start without any assumptions about the signature, and \emph{extract} the signature from other properties of the QG model itself. In this sense, this approach can be called an \emph{ab initio} approach to the question of the signature of the metric. Alternatively, turning the argument around, one can discuss what properties a given QG model ought to have in order to give a desired prediction for the signature of the metric (desired arguably being the Minkowski signature). This latter point of view is adopted in this paper --- we study the signature within one simple class of piecewise-flat QG models, and determine which detailed properties a model should have, in order to give an \emph{ab initio} prediction that the signature is of this or that type, Minkowski in particular.

To make things more concrete, we introduce a toy model of QG, similar to the Regge QG model, as follows. Given a spacetime manifold with some chosen topology, we specify its piecewise-flat structure by fixing a \emph{triangulation} (i.e., a simplicial complex), consisting of $4$-simplices, tetrahedra, triangles, edges and vertices. Then, one can introduce a path integral of the model, by integrating over all edge lengths in the triangulation:
\begin{equation}
Z = \cN \prod_m \int_D dl_m \, \exp (iS[l])\,.
\end{equation}
Here $\cN$ is an overall normalization constant, $m$ counts all edges in the triangulation, $D$ is the domain of integration for every edge (to be specified later), and $S[l]$ is the classical action of the theory, expressed as a functional over all edge lengths. For simplicity, we ignore the path integral over matter degrees of freedom, and discuss only gravity, since introducing matter fields does not change anything in our subsequent analysis. Given this definition of a path integral, one can evaluate the expectation value of an arbitrary observable $F(l)$ as follows:
\begin{equation} \label{eq:ToyModelQG}
\langle F \rangle = \frac{\cN}{Z} \prod_m \int_D dl_m \, F(l)\, \exp (iS[l])\,.
\end{equation}
The observable we are interested in is the frequency of observing the signature $s$ in the triangulation, denoted $P_s(l)$:
\begin{equation} \label{eq:observable}
P_s(l) = \frac{1}{|\sigma|} \sum_\sigma \chi_{\sigma,s}(l)\,, \qquad \text{ where } \qquad \chi_{\sigma,s}(l) = \left\{
\begin{array}{ccl}
1 & & \text{if signature}(\sigma) = s\,, \\
0 & & \text{if signature}(\sigma) \neq s\,. \\
\end{array}
\right.
\end{equation}
Here $\sigma$ denotes a $4$-simplex, $|\sigma|$ denotes the total number of $4$-simplices in the triangulation, and $\chi_{\sigma,s}(l)$ is an indicator function when the signature of $\sigma$ matches $s$. Substituting (\ref{eq:observable}) into (\ref{eq:ToyModelQG}) gives us the total probability of observing the signature $s$, as predicted by our toy model of QG. The main results of the paper are obtained by evaluating this probability numerically, using Monte Carlo method for numerical integration. Also, in order to simplify things, we completely omit the classical action, by setting it to zero, $S[l] = 0$, so that:
\begin{equation} \label{eq:ToyObservable}
\langle P_s \rangle = \frac{\cN}{Z} \prod_m \int_D dl_m \, P_s(l)\,.
\end{equation}
Therefore, we calculate the expectation value of $P_s(l)$ purely kinematically, by evaluating the path integral over all configurations of all edge lengths, ignoring the contributions coming from the dynamics of the theory (described by the exponent of the action above). Taking dynamics into account is out of the scope of this paper, and is left for future work. Note that the results for the probabilities $\langle P_s \rangle$ obtained in this paper possibly hold even for QG models other than the toy-model (\ref{eq:ToyModelQG}), as long as (\ref{eq:ToyObservable}) is an appropriate approximation of that model.

Aside from the actual question of the signature of the metric itself being interesting, the study of the signature, and the potential possibility of QG models to predict it, has additional motivation related to a research programme of generalizing the Coleman-Mandula (CM) theorem \cite{ColemanMandula}. Namely, as mentioned above, some QG models are built using the framework of higher gauge theory, describing the notion of a symmetry not by using an ordinary Lie group, but instead by using its categorical generalizations, called \emph{Lie $n$-groups}. Like for the case of supersymmetry, substituting a Lie group with a different mathematical structure naturally asks for a reexamination of the validity of the CM theorem. In the case of supersymmetry, it leads to the Haag-\L opusza\'{n}ski-Sohnius theorem \cite{HaagLopuszanskiSohnius}, while for the case of Lie $n$-groups the corresponding generalization is yet to be formulated. Within this research programme, one needs to discuss one of the main assumptions of CM theorem, namely that the Poincar\'e group is a subgroup of the full symmetry group of the theory. Since Poincar\'e group contains the Lorentz group $SO(3,1)$ as its subgroup, the validity of the above assumption can be questioned if the spacetime metric can have signature other than $(3,1,0)$. In other words, a QG model that is based on the notion of an $n$-group, and which treats the signature of the metric as a variable in the theory, represents an important framework in which the appropriate generalization of the CM theorem should be contextualized and studied. In this sense, understanding the implications of QG models for the signature of the metric represents one important step within the research programme related to generalizing the CM theorem.

The layout of the paper is as follows. In Section \ref{secII}, we introduce the theoretical concepts necessary to evaluate and discuss the signature of the spacetime metric. Section \ref{secIII} is devoted to a detailed description of the dataset generated and used for the numerical analysis. In Section \ref{secIV} we present the results of the numerical analysis. Finally, in Section \ref{secV} we give our concluding remarks and some topics for future research. The Appendix contains some additional technical results, and the tables with numerical data discussed in the main text.

\section{\label{secII}Theory}

We begin by reviewing the relationship between the edge lengths of a given $4$-simplex (and indeed, any $D$-simplex, where we will be primarily interested in the cases $D=2,3,4$) on one hand, and on the other hand, the signature of the metric induced by those edge lengths. As hinted in the Introduction, this relationship exists, and moreover, it is unique in the sense that any given set of edge lengths uniquely determines the whole metric tensor of the space in which the simplex can be naturally embedded, and thus its signature.

We establish this relationship in two independent ways --- one way relies on an explicit construction of the components of the metric tensor from edge lengths, in a suitable coordinate system. Finding its eigenvalues and counting the number of positive, negative and zero eigenvalues directly gives the signature of the metric. The other way relies on the notion of a Cayley-Menger determinant, which can be directly related to the metric signature, without actually constructing the components of the metric or evaluating its eigenvalues. The two methods of obtaining the signature of the metric are computationally sufficiently different that one can call them mutually independent. We employ both methods in order to check the consistency of our numerical calculations --- evaluating the signature using two different methods can signal a mistake if the results differ, and can reinforce the validity of the results if they coincide.

\subsection{The metric tensor of a $4$-simplex}

Given a $4$-simplex with vertices denoted as $v_0,\dots,v_4$, one can always embed it in a $4$-dimensional affine flat space, and use its edge lengths to introduce a metric in this space, which reproduces the required edge lengths when applied to the general formula for the squared distance between arbitrary points $A$ and $B$,
\begin{equation}
l_{AB}^2 = g_{\mu\nu} (x_A - x_B)^\mu (x_A - x_B)^\nu\,.
\end{equation}
This is obtained by integrating the differential geometry line element $ds^2 = g_{\mu\nu} dx^\mu dx^\nu$ along a straight line connecting the points $A$ and $B$, taking into account that the metric components $g_{\mu\nu}$ are constant along the line, since by assumption the embedding space is flat.

\begin{figure}[H]
\begin{center}
\begin{tikzpicture}
\coordinate (O)  at (-1, 0);
\coordinate (E0) at (-1.62, 1.8);
\coordinate (E1) at (0, 2.9);
\coordinate (E2) at (1.62, 1.8);
\coordinate (E3) at (1, 0);
\draw[thick] (O)  -- (E0) -- (E1) -- (E2) -- (E3) -- (O);
\draw[thick] (O)  -- (E1) ;
\draw[thick] (E1) -- (E3) ;
\draw[thick] (E3) -- (E0) ;
\draw[thick] (E0) -- (E2) ;
\draw[thick] (E2) -- (O) ;
\draw[very thin,-Stealth] ($(O)!1.3!(E0)$) -- ($(O)!2!(E0)$);
\draw[very thin,-Stealth] ($(O)!1.15!(E1)$) -- ($(O)!1.6!(E1)$);
\draw[very thin,-Stealth] ($(O)!1.15!(E2)$) -- ($(O)!1.6!(E2)$);
\draw[very thin,-Stealth] ($(O)!1.3!(E3)$) -- ($(O)!2!(E3)$);
\node[below left]  at (O)  {$v_4$};
\node[left] at (E0) {$v_0$};
\node[above] at (E1) {$v_1$};
\node[right] at (E2) {$v_2$};
\node[below right] at (E3) {$v_3$};
\node[right] at (-2.2,3.6) {$x^0$};
\node[right] at (0.56,4.65) {$x^1$};
\node[below] at (3.2,2.9) {$x^2$};
\node[below] at (3,0.05) {$x^3$};
\end{tikzpicture}
\end{center}
\caption{\label{fig:CoordinateSystem}Coordinate system associated to a $4$-simplex.}
\end{figure}
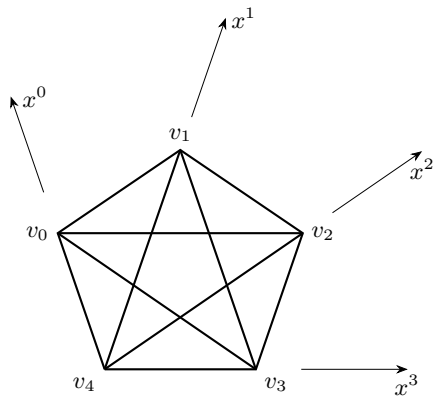

In order to obtain the components $g_{\mu\nu}$ explicitly in terms of edge lengths of the $4$-simplex, we need to specify a coordinate system in which the metric tensor has these components. A natural coordinate system that is well adapted to the geometry of the $4$-simplex is a rectilinear but not an orthogonal coordinate system. It is constructed by choosing one vertex of the $4$-simplex as a coordinate origin, and then by extending the four coordinate axes from the origin in the directions towards the remaining four vertices. See Figure \ref{fig:CoordinateSystem}.

Denoting the edge length between vertices $v_m$ and $v_n$ as $l_{mn}$, in this coordinate system the vertices of the $4$-simplex have the following coordinates:
\begin{equation} \label{eq:VertexCoords}
\begin{array}{rccccl}
v_0 = ( & |l_{40}|, & 0,  & 0, & 0 & )\,, \\ 
v_1= ( & 0, & |l_{41}|, & 0, & 0 & )\,, \\ 
v_2= ( & 0, & 0, & |l_{42}|, & 0 & )\,, \\ 
v_3= ( & 0, & 0, & 0, & |l_{43}| & )\,, \\ 
v_4= ( & 0, & 0, & 0, & 0 & )\,. \\ 
\end{array}
\end{equation}
The absolute values of the lengths ensure that coordinates are always real-valued, since the edge lengths themselves may be either real or imaginary. We can now write the equations for all edge lengths as
\begin{equation} \label{eq:edgesViaMetric}
l_{mn}^2 = g_{\mu\nu} (x_m - x_n)^\mu (x_m - x_n)^\nu\,, 
\end{equation}
where $x_m^\mu$ denotes the $\mu$-coordinate of the vertex $v_m$, as tabulated in (\ref{eq:VertexCoords}). Given that $m,n\in\{0,\dots,4\}$ and $\mu,\nu\in\{0,\dots,3\}$, equation (\ref{eq:edgesViaMetric}) represents a linear system of $15$ equations for $10$ unknowns $g_{\mu\nu}$ in terms of the $10$ known edge lengths $l_{mn}$. Five of these equations, corresponding to the choice $m=n$, take the form
\begin{equation}
l_{mm} = 0\,,
\end{equation}
reflecting the fact that the distance between a vertex and itself is always zero. The remaining ten equations, corresponding to the choice $m\neq n$, constitute a set of $10$ equations for the $10$ components of the metric tensor, and have a unique solution which can be written in the form:
\begin{equation} \label{eq:metricViaEdges}
g_{\mu\nu} = \frac{l_{4\mu}^2 + l_{4\nu}^2 - l_{\mu\nu}^2}{2 |l_{4\mu}| |l_{4\nu}|}\,.
\end{equation}
This provides us with the expression for the metric tensor as a function of the $10$ edge lengths of a $4$-simplex, in the coordinate system defined as in Figure \ref{fig:CoordinateSystem}. Note that the metric is always real and symmetric, regardless of whether the edge lengths are real or imaginary. From elementary linear algebra we know that real symmetric bilinear forms have real-valued eigenvalues, and each eigenvalue may be positive, negative or zero. One then defines the signature of the metric as a triple $(p,n,z)$, where $p$ is the number of positive eigenvalues, $n$ is the number of negative eigenvalues, and $z$ is the number of zero eigenvalues. Obviously, $p+n+z = 4$. Another fundamental theorem of linear algebra, Sylvester's law of inertia, guarantees that the signature of a bilinear form is independent of the choice of the basis used to express its components. Therefore, we can use the coordinate system introduced in Figure \ref{fig:CoordinateSystem}, evaluate the components of the metric using the formula (\ref{eq:metricViaEdges}), then use these components to evaluate the eigenvalues of the metric tensor, and finally count their signs to obtain the metric signature $(p,n,z)$. The final two steps, evaluating the eigenvalues and counting their signs, are of course done numerically.

We should note that the equation (\ref{eq:metricViaEdges}) can be generalized to an $D$-simplex living in a $D$-dimensional space, in a completely straightforward way. In that case, the vertex $v_D$ takes the role of the coordinate origin, while vertices $v_0,\dots,v_{D-1}$ are used to construct $D$ coordinate axes. Then, repeating the reasoning given above, one reaches the formula
\begin{equation} \label{eq:metricViaEdgesDdim}
g_{\mu\nu} = \frac{l_{D\mu}^2 + l_{D\nu}^2 - l_{\mu\nu}^2}{2 |l_{D\mu}| |l_{D\nu}|}\,, \qquad \mu,\nu\in \{0,\dots,D-1\} \,.
\end{equation}
In the remainder of this paper, in addition to the case $D = 4$ which provides the metric of a $4$-simplex, we will also employ the cases $D = 3$ and $D = 2$ to evaluate the signature of the metric tensor associated to a tetrahedron and to a triangle, respectively.

Finally, let us also note that the metric tensor (\ref{eq:metricViaEdgesDdim}) is sometimes called the \emph{normalized Gram matrix} in the literature, whereas the ordinary, non-normalized Gram matrix is usually introduced as the numerator of (\ref{eq:metricViaEdgesDdim}), and denoted as $G_{\mu\nu} =  l_{D\mu}^2 + l_{D\nu}^2 - l_{\mu\nu}^2$.  The main difference is that the Gram matrix is a dimensionful quantity, having the dimension of squared length, while the metric tensor is dimensionless, as expected from the definition of distance (\ref{eq:edgesViaMetric}).

\subsection{\label{subsec:CayleyMenger}The Cayley-Menger determinants of a $4$-simplex}

An alternative, purely algebraic way of evaluating the signature of the metric in a $4$-simplex is based on Cayley-Menger determinants. Given a $D$-simplex, denoting its vertices as $v_0,\dots,v_{D}$ and the length of an edge between vertices $v_i$ and $v_j$ as $l_{ij}$, the Cayley-Menger determinant for that $D$-simplex, denoted $C_D$, is defined as:
\begin{equation} \label{eq:CMdetDef}
C_D = \det \left[
\begin{array}{ccccccc}
 0 & 1 & 1 & 1 &\cdots & 1 \strut \\
 1 & 0 & l_{01}^2  & l_{02}^2 & \cdots & l_{0D}^2 \strut \\
 1 & l_{01}^2 & 0  & l_{12}^2 & \cdots & l_{1D}^2 \strut \\
 1 & l_{02}^2 & l_{12}^2 & 0 & \cdots & l_{2D}^2 \strut \\
 \vdots & \vdots & \vdots & \vdots & \ddots & \vdots \strut \\
 1 & l_{0D}^2 & l_{1D}^2 & l_{2D}^2 & \cdots & 0 \strut \\
\end{array}
\right]\,.
\end{equation}
The matrix in (\ref{eq:CMdetDef}) is a symmetric matrix of dimension $(D+2)\times (D+2)$, containing the squares of all edge lengths of the $D$-simplex. Therefore, since the square of an edge length is always real-valued, the matrix is also real, and consequently the Cayley-Menger determinant itself. For $D=0$ we have $C_0 = -1$, for $D=1$ we have $C_1 = 2 l_{01}^2$, and so on. The square of the $D$-volume of a $D$-simplex, $V_D$, can be evaluated in terms of the squares of its edge lengths as 
\begin{equation}
V_D^2 = \frac{(-1)^{D+1}}{2^D \, (D!)^2} C_D\,.
\end{equation}
This is a generalization of the Heron's formula for the area of a triangle, Tartaglia's formula for the volume of a tetrahedron, and so on. It enables one to establish a relation between the Cayley-Menger determinant and the determinant of the metric tensor (\ref{eq:metricViaEdgesDdim}), as follows:
\begin{equation} \label{eq:relationCMandMetricDet}
C_D = (-1)^{D+1} 2^D \Big( \prod_{m=0}^{D-1} | l_{Dm}^2 | \Big) \det g_{\mu\nu}\,.
\end{equation}
This relationship gives rise to the following theorem, that allows one to evaluate the signature of the metric by evaluating the Cayley-Menger determinants of the $4$-simplex and its subsimplices.

\textbf{Theorem.}
\textit{Given a $D$-simplex with vertices denoted as $v_0,\dots,v_D$, introduce a sequence of $m$-simplices $\sigma_m$ constructed as subsimplices of the $D$-simplex, spanned over the following vertices:
\begin{equation}
\sigma_0 = \sspan(v_D)\,, \quad \sigma_1 = \sspan(v_{D-1},v_D)\,, \quad \dots \quad \sigma_m = \sspan(v_{D-m},\dots,v_D)\,, \quad \dots\quad \sigma_D = \sspan(v_0,\dots,v_D)\,.
\end{equation}
Assume that all corresponding Cayley-Menger determinants are nondegenerate, $C_m \neq 0$ for $m\in\{ 0,\dots,D \}$, and construct the sets
\begin{equation}
P = \Big\{  m\in\{1,\dots,D\}  \, \Big| \,  \frac{C_m}{C_{m-1}} < 0 \,  \Big\}\,, \qquad
N = \Big\{  m\in\{1,\dots,D\}  \, \Big| \,  \frac{C_m}{C_{m-1}} > 0 \,  \Big\}\,.
\end{equation}
Denote the numbers of elements of sets $P$ and $N$ as $|P|$ and $|N|$, respectively. Then, the signature of the metric corresponding to the $D$-simplex $\sigma_D$ is given by
\begin{equation} \label{eq:SignatureFromCM}
(p,n,z) = (|P|,|N|,0)\,.
\end{equation}
In other words, $p$ and $n$ are the \emph{number of alterations} and the \emph{number of preservations} of the signs of the following sequence Cayley-Menger determinants:
\begin{equation}
C_0 \equiv -1, \quad C_1\,, \quad C_2\,, \quad \dots\quad C_{D-1}\,,\qquad C_D\,.
\end{equation}
}
The proof of the Theorem is given in Appendix \ref{AppA}.

As we can see, the above Theorem enables us to determine the signature of the metric associated to a $D$-simplex by evaluating the appropriate sequence of Cayley-Menger determinants, and counting the number of sign changes in that sequence. This method relies on the prior assumption that the metric is nondegenerate, which is reflected in the assumption that all determinants $C_k$ are nonzero, and consequently $z=0$ in the signature (\ref{eq:SignatureFromCM}). Therefore, the Cayley-Menger method is not as general as the direct evaluation of eigenvalues of (\ref{eq:metricViaEdgesDdim}), but its advantage is that it is fully algebraic, without requiring one to find $D$ roots of the characteristic polynomial of (\ref{eq:metricViaEdgesDdim}).

For the purpose of the numerical analysis that is performed in the rest of this work, the requirement of nondegeneracy is not a practical limitation, since all simplices we discuss turn out to be nondegenerate. In this sense, the method of evaluating the signature of the metric established by the above Theorem can be regarded as a viable alternative to the method of evaluating eigenvalues of (\ref{eq:metricViaEdgesDdim}). Therefore, in our numerical analysis, we apply both methods simultaneously, as a consistency check of the validity of the obtained metric signatures.

\subsection{\label{subsec:LengthScales}Length scales of the triangulation}

In any QG model based on piecewise-flat spacetime, there are at least two relevant physical length scales that describe the piecewise-flat structure --- the \emph{minimal} and \emph{maximal} length of any edge in any particular $4$-dimensional cell in the lattice, denoted $l_\text{min}$ and $l_\text{max}$.

The minimal length scale has the physical interpretation of a UV cutoff. Its value is in the domain $0 \leq l_\text{min} < l_\text{max}$, and the value of zero corresponds to a model with no UV cutoff. Most commonly, the minimal length scale is identified with the Planck scale, $l_p = 10^{-35}\,{\rm m}$, although this is merely a guess, since we have no experimental evidence for the actual value of $l_\text{min}$. On the other hand, the maximal length scale, estimating the maximum ``size'' of lattice cells, represents the scale beyond which the spacetime manifold cannot be considered smooth, i.e., for which its piecewise-flat structure becomes apparent. Current experimental evidence places an upper bound on this scale. Namely, we observe that the spacetime manifold can be considered smooth in a huge range of scales --- from the scale of the observable Universe, approximately $10^{26}\,{\rm m}$, down to the scale probed by current experiments at the LHC and at LIGO/Virgo observatories, which is roughly at $10^{-20}\,{\rm m}$. We therefore have the following rough phenomenological estimate:
\begin{equation} \label{eq:phenomenology}
0 \leq l_\text{min} < l_\text{max} < 10^{-20} \,{\rm m}\,, \qquad \text{ maybe with } \qquad l_\text{min} \simeq l_p = 10^{-35}\,{\rm m}\,.
\end{equation}

The metric tensor, by its definition, is a scale-invariant quantity, in the sense that if we rescale all edge lengths $l_n$ (where index $n$ counts all edges in the triangulation) by a common factor $\lambda \in \realni^+$,
\begin{equation} \label{eq:rescalingSymmetry}
l_n \quad \to \quad l_n' = \lambda l_n\,,
\end{equation}
the metric tensor will remain the same, as per (\ref{eq:metricViaEdgesDdim}). This property can be exploited when calculating the signature of the metric, which is also scale-invariant. Namely, the equation (\ref{eq:metricViaEdgesDdim}) corresponds to the components of the metric for a single $D$-dimensional cell, and holds only for the edge lengths belonging to the intervals
\begin{equation}
l_n \in (l_\text{min},l_\text{max}]\,, \qquad \text{ or }\qquad l_n \in (il_\text{min},il_\text{max}]\,,
\end{equation}
respectively for spacelike and timelike edges. Technically, the intervals are open on the left-hand side since in (\ref{eq:metricViaEdgesDdim}) one has to exclude the value $l_\text{min} = 0$, although one is allowed to choose any length arbitrarily close to $l_\text{min}$. Given these domains, we can exploit the rescaling symmetry (\ref{eq:rescalingSymmetry}) to pass from actual lengths (measured in, say, meters) to \emph{unitless} lengths, via the substitution
\begin{equation}
l_n \quad \to \quad l_n^\text{unitless} = 10 \frac{l_n}{l_\text{max}}\,,
\end{equation}
where the factor of $10$ is introduced purely for the convenience of numerical computations, discussed in the rest of the paper. The new unitless lengths belong to the intervals
\begin{equation} \label{eq:unitlessLengthDomains}
l_n^\text{unitless} \in ( k ,10]\,, \qquad \text{ or }\qquad l_n^\text{unitless} \in ( ki ,10i]\,,
\end{equation}
depending on their spacelike or timelike nature. Here, the unitless parameter $k$ is given as
\begin{equation}
k \equiv 10\frac{l_\text{min}}{l_\text{max}}\,, \qquad 0\leq k < 10\,,
\end{equation}
and it now assumes the role of the UV cutoff (from now on, simply called \emph{cutoff}). Determining the value of $k$ gives us insight into the ratio of scales $l_\text{min}$ and $l_\text{max}$. For example, assuming that $l_\text{min} = l_p$, we have
\begin{equation} \label{eq:translationForLmax}
l_\text{max} = \frac{10}{k} l_p\,,
\end{equation}
which tells us the maximum possible size of an individual cell in a triangulation, in terms of the Planck length. The phenomenological bounds (\ref{eq:phenomenology}) then translate into the range $10^{-14} < k < 10$. Therefore, knowing the value of $k$ places a restriction on any QG model based on piecewise-flat spacetime, since every model must ensure that the lengths of all edges are always below the value of $l_\text{max}$.

\section{\label{secIII}Description of the dataset}

After establishing the procedure, in fact two independent procedures, to evaluate the signature of the metric in a given $4$-simplex from the information about its edge lengths, the basic idea is to choose some random triangulation, assign each edge a random length, evaluate the signature at each $4$-simplex, and count the frequencies of appearance of distinct signatures. While this is an elementary underlying idea, implementing it in practice requires dealing with many additional details.

In order to fully specify all details of our numerical analysis, we first focus on the description and the construction of the dataset that represents the input for the numerical calculations. We describe the construction of the full dataset in stages, as follows.

\subsection{Fixing the topology}
  
In order to investigate whether, and how much, our results depend on the choice of the topology of the underlying spacetime manifold, we construct the dataset for three different topologies:
\begin{enumerate}
\item a $4$-ball, with a $3$-sphere as its boundary,
\item a $4$-sphere, with no boundary, and
\item a disconnected set of $4$-simplices, each with its five tetrahedra as the boundary. 
\end{enumerate}
Evaluating the signature of the metric in these three distinct cases enables us to quantify how much the result depends on the connectedness between $4$-simplices, i.e., how much the frequencies of different signatures change depending on whether the $4$-simplices are connected to each other along their common tetrahedra, or not. In addition, the difference between a $4$-ball and a $4$-sphere will allow us to quantify the dependence of the signature on the existence of the boundary in a manifold.

\subsection{Choosing the triangulation}

Given each of the three topologies, we construct \emph{random triangulations} that serve as a basis for the calculation of the signature. In order to quantify how the frequencies of the signatures depend on the choice of a triangulation itself, for the topologies of a $4$-ball and a $4$-sphere we construct $10$ different random triangulations, so that their combinatorial structure (what is adjacent to what) is different in each of the $10$ cases. For the case of the disconnected topology, constructing a triangulation is trivial, since there exists only one possible choice --- a set of mutually non-adjacent $4$-simplices. Thus, we will work with $2\times 10 + 1 = 21$ overall distinct triangulations.

Each triangulation contains approximately $10^5$ $4$-simplices (up to irrelevant fluctuations of $\pm 4$ in their total number). Overall, this amounts to $2.1$ million $4$-simplices in the dataset.

In the case of a $4$-sphere, the triangulations are being constructed as follows. We start from an initial triangulation of a $4$-sphere as a boundary of a $5$-simplex, made up of $6$ interconnected $4$-simplices. We then randomly apply $1\to 5$ and $2\to 4$ Pachner moves to randomly chosen $4$-simplices and tetrahedra, respectively, whenever the moves are applicable. This is being iterated until the total number of $4$-simplices in the triangulation becomes greater or equal to $10^5$, giving us the desired triangulation. The random number generator implements a uniform probability distribution over the current pool of $4$-simplices and tetrahedra at each step. In this way we generate $10$ differerent triangulations, where the random number generator essentially guarantees that each triangulation will have different combinatorial structure from the others.

In the case of the $4$-ball topology, we apply the same construction as for the case of a $4$-sphere, albeit with few important additions. First, the initial triangulation is chosen to be a single $4$-simplex with $5$ boundary tetrahedra. From there, the construction of the triangulation proceeds by growing the boundary, followed by growing the bulk. The boundary is grown by randomly choosing a boundary tetrahedron, and attaching to it a new $4$-simplex, making the tetrahedron and the new $4$-simplex part of the bulk, and making the remaining $4$ tetrahedra of the new $4$-simplex part of the new boundary. This procedure is iterated until a desired size of the boundary (i.e., the desired number of boundary tetrahedra) is reached. Then, the bulk is grown using the Pachner moves, in the same way as for the $4$-sphere.

In order to implement this construction, we need to specify the desired bulk-to-boundary ratio of the resulting triangulation. Namely, a $4$-ball can be triangulated so that it has very few boundary tetrahedra, compared to the number of bulk $4$-simplices. This leads to a ``shape'' of a very dense packing of $4$-simplices with a minimal ball-shaped boundary. On the other hand, we can choose to have many boundary tetrahedra, compared to a comparatively few $4$-simplices in the bulk. This leads to a ``shape'' that resembles a snowflake-like structure. In order to avoid either extreme case, we proceed as follows. In ordinary Euclidean geometry, the total $4$-volume and a total $3$-area of a $4$-ball depend on the radius of the ball as
\begin{equation}
V_4 = \frac{1}{2} \pi^2 r^4\,, \qquad A_3 = 2\pi^2 r^3\,.
\end{equation}
Eliminating the radius, we obtain the relation
\begin{equation}
A_3 = \sqrt[4]{128 \pi^2}\, V_4^\frac{3}{4} \,.
\end{equation}
The numerical coefficient is $\sqrt[4]{128 \pi^2} \approx 6 $, so if the $4$-volume consists of the desired $10^5$ $4$-simplices, we obtain that the $3$-area should consist of roughly $33\,740$ boundary tetrahedra. Therefore, if we round off the number of boundary tetrahedra to $3\,500$ and grow the boundary to that number, and then grow the bulk to the number $10^5$ $4$-simplices, the resulting triangulation will have the bulk-to-boundary proportion roughly corresponding to a ``shape'' of a regular $4$-ball, which is arguably ``well balanced'' compared to the two extreme configurations discussed above. Of course, strictly speaking, the ratio of $35\,000$ boundary tetrahedra to $10^5$ $4$-simplices is actually an ad-hoc choice, and the naive reasoning about the various ``shapes'' of the triangulation is not actually well defined. Nevertheless, this naive reasoning merely reflects our intuition from Euclidean space, and lacking any better criteria, we decide to adopt the above choice of bulk-to-boundary proportion.

\subsection{Domains of the edge lengths}

Given each of the $21$ constructed triangulations, the next task in the specification of the dataset is to assign a length to each edge in the triangulation. There are three qualitatively distinct possible choices:
\begin{enumerate}
\item all edges are spacelike,
\item all edges are timelike, and
\item edges are mixed, some being spacelike and others timelike.
\end{enumerate}
Note that lightlike edge lengths have magnitude zero, and can give rise to degenerate metrics and Cayley-Menger determinants, so we exclude the lightlike case. Keeping in mind the scale-free nature of the metric, as discussed in Subsection \ref{subsec:LengthScales}, we work with the unitless edges (from now on ``unitless'' is implicit when we talk about edge lengths), with the domains given by (\ref{eq:unitlessLengthDomains}). Therefore, in the case of the spacelike edges, we have
\begin{equation}
l_n \in ( k ,10]\,,
\end{equation}
in the case of the timelike edges we instead have
\begin{equation}
l_n \in ( ki ,10i]\,,
\end{equation}
while in the case of the mixed edges we have
\begin{equation}
l_n \in (k,10] \cup (ki,10i]\,.
\end{equation}
The lengths are randomly chosen from these sets via the uniform distribution. In the case of mixed edges, this is equivalent to choosing each edge as timelike or spacelike via a fair coin toss, after which its length is sampled from the appropriate interval. 

The cutoff parameter $k$ in general satisfies $0 \leq k < 10$. In order to facilitate various sampling choices of $k$ which are reasonable for a numerical analysis, we opt to restrict the domain of the cutoff to a set of integer values $k \in \{ 0,\dots,9 \}$. This allows us to quantify how the frequencies of the signatures depend on the existence and the value of the cutoff. With this choice, the spacelike, timelike, and mixed edge lengths are each chosen from a set of $10$ possible domains, one for each choice of $k$. Therefore, in total there are $10 + 10 + 10 = 30$ possible choices for the domain of the edge lengths, for each of the $21$ triangulations corresponding to three topologies. That makes up a total of $30 \times 21 = 630$ distinct datasets.

\subsection{Fluctuations in the choices of edge lengths}

Given a particular topology, and a particular triangulation of that topology, and a particular domain for the choice of edge lengths, it is not enough to assign lengths to edges \emph{only once}. Namely, we want to quantify how the frequencies of the signatures of the metric fluctuate as we choose \emph{several different assignments} od lengths to edges, for the same triangulation and domain of lengths. To that end, for each choice of the topology, triangulation and domain for the edges, we independently assign lengths to edges in $10$ different \emph{runs}, constructing $10$ different datasets in each case. Comparing the frequencies of signatures across multiple runs allows us to quantify how much they can vary, on a fixed triangulation and for the fixed domain of edge lengths. Given that we have a total of $630$ choices of these, constructing $10$ runs for each amounts to a grand total of $6\,300$ datasets.

\subsection{The total dataset}

This concludes the detailed specification of the dataset structure. Let us also note the full size of the dataset, and some details about generating and storing the data. Each of the $6\,300$ individual datasets contains approximately $10^5$ $4$-simplices, and for each $4$-simplex and its corresponding $5$ tetrahedra and $10$ triangles, we evaluate the signature of the metric, via two independent methods. This amounts to a total pool of
\begin{equation}
  6\,300 \times 10^5 \times (1+5+10) \times 2 =2.016 \cdot 10^{10}
\end{equation}
evaluations of the signature of the metric of various $4$-, $3$-, and $2$-simplices, i.e., slightly over $20$ billion of them, performed for the purpose of our work. The results are collected in $6\,300$ independent \texttt{.csv} files, each file containing approximately $10^5$ rows and $33$ columns, containing all evaluated data (Cayley-Menger determinant and the signature evaluated diretly from the metric, for each of the $1+5+10$ simplices in $32$ columns) and all particular assignments of edge lengths (in the $33$-rd column of the file). All resulting \texttt{.csv} files, and the algorithms developed in \texttt{C++} and \texttt{Wolfram Mathematica} used to generate the \texttt{.csv} files, are publicly available (open access) at \cite{Zenodo1,Zenodo2,Zenodo3,Zenodo4}.

In the next Section we turn to the analysis of the obtained results.

\section{\label{secIV}Analysis of numerical results}

After generating the dataset described in Section \ref{secIII}, and performing appropriate statistical analysis over this dataset, in this Section we report the results and draw a number of conclusions. Our central point of interest are the frequencies of appearence of the five possible nondegenerate signatures in $4$-dimensional spacetime, namely:
\begin{itemize}
\item the Euclidean signature $(4,0,0)$,
\item the Minkowski signature $(3,1,0)$,
\item the split signature $(2,2,0)$,
\item the negative Minkowski signature $(1,3,0)$, and
\item the negative Euclidean signature $(0,4,0)$.
\end{itemize}
As a curiosity, let us note here that the split signature is also called Kleinian, neutral, ultrahyperbolic, and Atiyah's signature throughout the literature, depending on the author. We opt to call it \emph{split} because we feel this term is both short and descriptive.

In what follows, the main analysis and discussion revolves around the results corresponding to the topology of a disconnected set of $4$-simplices. As it turns out, all main conclusions can be drawn from the results obtained for the disconnected topology, so we postpone the discussion of the $4$-ball and the $4$-sphere topologies for later. The parameters of the (trivial) triangulation corresponding to the disconnected topology are given in Table \ref{tab:parametersDisconnectedTopology}.
\begin{table}[H]
\begin{center}
\begin{tabular}{|c|c|c|c|c|c|} \hline
\ disconnected topology\ \ &\ vertices\ \ &\ edges\ \ &\ triangles\ \ &\ tetrahedra\ \ &\ $4$-simplices\ \ \\ \hline\hline
1 &\ 500\,000\ &\  1\,000\,000\ &\ 1\,000\,000\ &\ 500\,000\ &\ 100\,000\ \ \\ \hline
\end{tabular}
\end{center}
\caption{\label{tab:parametersDisconnectedTopology}Parameters of the single triangulation corresponding to the disconnected topology. The numbers simply reflect the fact each simplex has $5$ vertices, $10$ edges, $10$ triangles and $5$ tetrahedra.}
\end{table}

\subsection{Spacelike edge domain}

\begin{figure}[H]
\begin{center}
\includegraphics[scale=0.5]{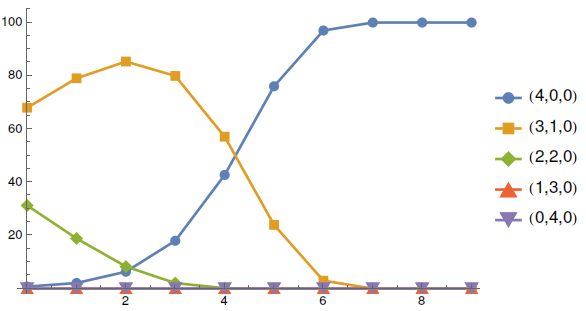}
\end{center}
\caption{\label{fig:tab1}Frequencies of various signatures, for the disconnected topology with spacelike domain for edge lengths. For each signature, and for each value of the cutoff $k$ on the horizontal axis, we plot the mean frequency as a percentage (\%). The corresponding sample standard deviations satisfy $\sigma \leq 0.2\,\%$ and are thus too small to be visible on the plot. The mean is obtained by averaging over $10$ sample runs, assigning a different set of edge lengths in each run. The data for this graph is taken from Table \ref{tab1}.}
\end{figure}

Let us begin by focusing on the choice of spacelike domain for edge lengths. After averaging over $10$ sample runs, the frequencies of all five signatures are expressed as percentages relative to a total number of $10^5$ $4$-simplices, along with the corresponding sample standard deviations $\sigma$, also expressed as percentages. The dependence of the frequencies on the choice of the cutoff $k\in\{0,\dots,9\}$ is presented in Figure \ref{fig:tab1}.

By far the most important feature on Figure \ref{fig:tab1} is the dominance of the Minkowski signature for cutoff values in the interval $k\in [0,4]$. The maximum is achieved in the neighborhood of $k=2$, and looking at the corresponding values in Table \ref{tab1} in Appendix \ref{AppB}, we see the relative frequencies of the five signatures:
\begin{equation}
\begin{array}{lcr}
(3,1,0): && 85.34 \pm 0.07\, \%\,, \\
(2,2,0): && 8.30 \pm 0.06\, \%\,, \\
(4,0,0): && 6.36 \pm 0.06\, \%\,, \\
(1,3,0): && 0 \, \% \,, \\
(0,4,0): && 0 \, \%\,. \\
\end{array}
\end{equation}
This represents the \emph{first main result} of the paper. Namely, in light of the discussion at the end of Subsection \ref{subsec:LengthScales}, if we have a QG model with a UV cutoff $l_\text{min}$ at the Planck scale, and if the model implements a path integral over spacelike edge lengths with the domain for each edge given by
\begin{equation} \label{eq:domainForMinkowskiSig}
l_n \in [ l_p, 5 l_p]\,,
\end{equation}
such a model generically \emph{predicts the probability of observing the Minkowski signature at any point in spacetime} to~be~$85 \,\%$. The remaining $15\,\%$ are roughly equally split between the Euclidean signature and the split signature, while the two negative signatures are completely absent. Moreover, these probabilities are based only on kinematics, i.e., on the properties of the configuration space for edge lengths based on (\ref{eq:domainForMinkowskiSig}). Introducing the dynamics into the calculation may change this probability, with a possibly even higher probability for Minkowski signature. Of course, dynamics can also spoil this result by reducing the probability. The influence of dynamics is model dependent, and is a topic for future work. Nevertheless, even looking at the kinematics only, the above result is arguably significant.

Of course, from Figure \ref{fig:tab1} we can also see that other values of the cutoff lead to different results, but this is expected and easy to explain. Namely, the dominance of the Euclidean metric for $k\in[6,10]$ can be translated via (\ref{eq:translationForLmax}) to the edge lengths from the domains such as
\begin{equation} \label{eq:domainForEuclideanSig}
k=6: \quad l_n \in [ l_p, 1.6\, l_p]\,, \qquad \text{ to } \qquad k=9: \quad l_n \in [ l_p, 1.1\, l_p]\,.
\end{equation}
As opposed to (\ref{eq:domainForMinkowskiSig}), these domains are very narrow (relative to $l_p$ itself), implying that all edge lengths must be largely concentrated near a single value. This gives rise to mostly equilateral $4$-simplices, which are of course embeddable only in Euclidean spacetime, leading to the dominance of the Euclidean signature for large values of the cutoff $k$.

Regarding the split signature, Figure \ref{fig:tab1} indicates that its contribution starts at $31.2\,\%$ for $k=0$, and steadily drops afterwards, becoming completely irrelevant already around $k=4$. The split signature is interesting from a theoretical point of view, since it describes a spacetime with two space and two time axes. However, its contribution is never dominant, and hardly even substantial in the case of spacelike edge domain.

The negative Euclidean signature has strictly zero probability in Figure \ref{fig:tab1}. This is also easily explainable, and is in fact a theorem. Namely, in a Cartesian coordinate system, with coordinates denoted as $x^\mu$ (not to be confused with the non-Cartesian coordinate system of Figure \ref{fig:CoordinateSystem}), the formula for evaluating the distance between spacetime points $A$ and $B$ that corresponds to the metric with negative Euclidean signature reads:
\begin{equation}
l_{AB} = \sqrt{ - (x^0_A - x^0_B)^2 - (x^1_A - x^1_B)^2- (x^2_A - x^2_B)^2- (x^3_A - x^3_B)^2 }\,.
\end{equation}
Obviously, since the coordinates are always real-valued, all distances evaluated using this metric must be purely imaginary, i.e., timelike. Therefore, since by assumption we work in the purely spacelike edge domain, a $4$-simplex with inverse Euclidean metric signature is impossible to construct, giving exactly $0\,\%$ probability for this signature.

Finally, the negative Minkowski signature, although practically invisible in Figure \ref{fig:tab1}, has a tiny, yet nonzero probability, which is a largely counter-intuitive result. For the cutoff $k=0$, Table \ref{tab1} gives the frequency $0.003 \, \% \pm 0.002 \,\%$. In our total pool of $2.1 \cdot 10^8$ constructed $4$-simplices with spacelike edge lengths within the full dataset, we have found a total of $507$ $4$-simplices with signature $(1,3,0)$. Aside from being extremely rare, the only interesting property of the negative Minkowski signature is that it describes a spacetime with one space and three time axes. Table \ref{tab:negativeMinkowskiExamples} lists the choices of edge lengths for five example $4$-simplices with negative Minkowski signature.
\begin{table}[H]
\begin{center}
  \begin{tabular}{|c|c|c|c|c|c|c|c|c|c|c|} \hline
\ $4$-simplex\ \ &\ $l_{01}$\ \ &\ $l_{02}$\ \ &\ $l_{03}$\ \ &\ $l_{04}$\ \ &\ $l_{12}$\ \ &\ $l_{13}$\ \ &\ $l_{14}$\ \ &\ $l_{23}$\ \ &\ $l_{24}$\ \ &\ $l_{34}$\ \  \\ \hline\hline
1\ \ &\ 9.36\ \ &\ 1.18\ \ &\ 5.96\ \ &\  3.76\ \ &\ 2.06\ \ &\ 0.91\ \ &\ 1.59\ \ &\ 1.01\ \ &\ 0.09\ \ &\ 0.39\ \  \\ \hline
2\ \ &\ 3.08\ \ &\  0.14\ \ &\  3.31\ \ &\  0.67\ \ &\  0.50\ \ &\  9.71\ \ &\  7.65\ \ &\  4.94\ \ &\  2.82\ \ &\  0.68\ \  \\ \hline
3\ \ &\ 5.04\ \ &\  0.18\ \ &\  2.19\ \ &\  8.06\ \ &\  6.70\ \ &\  0.42\ \ &\  2.91\ \ &\  4.37\ \ &\  9.79\ \ &\  4.66\ \  \\ \hline
4\ \ &\ 3.96\ \ &\  2.24\ \ &\  0.25\ \ &\  0.74\ \ &\  0.61\ \ &\  1.91\ \ &\  6.99\ \ &\  0.37\ \ &\  4.20\ \ &\  2.31\ \  \\ \hline
5\ \ &\ 0.11\ \ &\  6.69\ \ &\  3.27\ \ &\  0.15\ \ &\  3.97\ \ &\  0.82\ \ &\  1.37\ \ &\  1.43\ \ &\  8.41\ \ &\  4.87\ \ \\ \hline
\end{tabular}
\end{center}
\caption{\label{tab:negativeMinkowskiExamples}Values of edge lengths for the five example $4$-simplices with negative Minkowski signature. The vertices of the $4$-simplex are labeled as in Figure \ref{fig:CoordinateSystem}.}
\end{table}

\subsection{Timelike edge domain}

\begin{figure}[H]
\begin{center}
\includegraphics[scale=0.5]{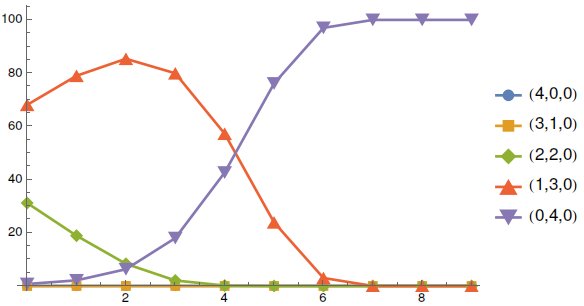}
\end{center}
\caption{\label{fig:tab2}Frequencies of various signatures, for the disconnected topology with timelike domain for edge lengths. For each signature, and for each value of the cutoff $k$ on the horizontal axis, we plot the mean frequency as a percentage (\%). The corresponding sample standard deviations satisfy $\sigma \leq 0.2\,\%$ and are thus too small to be visible on the plot. The mean is obtained by averaging over $10$ sample runs, assigning a different set of edge lengths in each run. The data for this graph is taken from Table \ref{tab2}.}
\end{figure}

Next, we turn to the choice of timelike domain for edge lengths. Similarly to the spacelike case, after averaging over $10$ sample runs, the frequencies of all five signatures are expressed as percentages relative to a total number of $10^5$ $4$-simplices, along with the corresponding sample standard deviations $\sigma$, also expressed as percentages. The dependence of the frequencies on the choice of the cutoff $k\in\{0,\dots,9\}$ is presented in Figure \ref{fig:tab2}.

This time, by far the most prominent feature on Figure \ref{fig:tab2} is that it looks identical to its spacelike counterpart \ref{fig:tab1}. Essentially, the only difference is that the signs of the signatures are now reversed:
\begin{itemize}
\item the cutoff range $k\in[0,4]$ is dominated by negative Minkowski signature $(1,3,0)$,
\item the cutoff range $k\in[6,10]$ is dominated by negative Euclidean signature $(0,4,0)$,
\item the split signature displays behavior identical to the spacelike case,
\item the Minkowski signature is now extremely rare, with relative frequency $0.002 \,\% \pm 0.002\,\%$ for $k=0$,
\item the Euclidean signature has strictly zero probability.
\end{itemize}

These results represent a sort of a mirror image of the results obtained for the spacelike case. And indeed, this is not an accident. Namely, for any spacelike edge length, evaluated using some metric $g_{\mu\nu}$ and some coordinates of spacetime points $A$ and $B$,
\begin{equation} \label{eq:spacelikeEdgeEval}
l_{AB}^2 = g_{\mu\nu} (x_A - x_B)^\mu (x_A - x_B)^\nu\,,
\end{equation}
one can introduce a corresponding timelike edge and the metric with the opposite sign,
\begin{equation} \label{eq:timelikeSpacelikeIsomorphism}
\tilde{l}_{AB} = i l_{AB}\,, \qquad \tilde{g}_{\mu\nu} = - g_{\mu\nu}\,,
\end{equation}
such that the timelike edge length can be evaluated using the new metric and the same coordinates of spacetime points $A$ and $B$:
\begin{equation} \label{eq:timelikeEdgeEval}
\tilde{l}_{AB}^2 = \tilde{g}_{\mu\nu} (x_A - x_B)^\mu (x_A - x_B)^\nu\,.
\end{equation}
Substituting (\ref{eq:timelikeSpacelikeIsomorphism}) into (\ref{eq:timelikeEdgeEval}) obviously recovers (\ref{eq:spacelikeEdgeEval}).

In other words, given any triangulation of spacetime, and any purely spacelike set of edge lengths assigned to that triangulation, one can always assign a purely timelike set of edge lengths to the same triangulation, with the only difference being the flip in the sign of the metric tensor. There is a \emph{complete isomorphism} between purely spacelike and purely timelike configurations of edge lengths, and any QG model featuring a path integral over the spacelike edge length domain can be substituted with a fully equivalent QG model featuring a path integral over the timelike edge length domain, provided that one also makes a substitution $g_{\mu\nu} \to - g_{\mu\nu}$ everywhere in the model.

Therefore, choosing between the purely spacelike and purely timelike domain for edge lengths is simply a matter of convention, and either can be used in an equivalent fashion. As a consequence, all results discussed in the previous Subsection hold here as well, albeit with the signs of all signatures flipped.

\subsection{Mixed edge domain}

\begin{figure}[H]
\begin{center}
\includegraphics[scale=0.5]{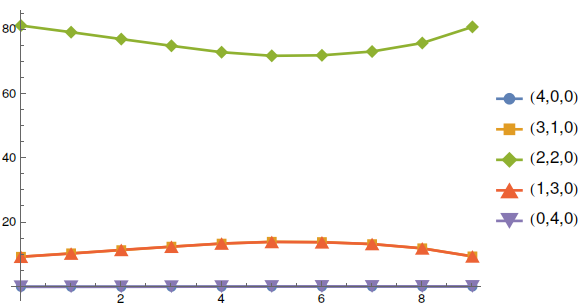}
\end{center}
\caption{\label{fig:tab3}Frequencies of various signatures, for the disconnected topology with mixed domain for edge lengths. For each signature, and for each value of the cutoff $k$ on the horizontal axis, we plot the mean frequency as a percentage (\%). The corresponding sample standard deviations satisfy $\sigma \leq 0.2\,\%$ and are thus too small to be visible on the plot. The mean is obtained by averaging over $10$ sample runs, assigning a different set of edge lengths in each run. The data for this graph is taken from Table \ref{tab3}.}
\end{figure}

The next step is the choice of mixed domain for edge lengths. Similarly to the spacelike and timelike cases, after averaging over $10$ sample runs, the frequencies of all five signatures are expressed as percentages relative to a total number of $10^5$ $4$-simplices, along with the corresponding sample standard deviations $\sigma$, also expressed as percentages. The dependence of the frequencies on the choice of the cutoff $k\in\{0,\dots,9\}$ is presented in Figure \ref{fig:tab3}.

In this case, the most prominent feature in Figure \ref{fig:tab3} is the complete dominance of the split signature, for all cutoff values. The maximum is achieved in the neighborhood of $k=0$, and looking at the corresponding values in Table \ref{tab3} in Appendix \ref{AppB}, we see the relative frequencies of the five signatures for $k=0$ are:
\begin{equation}
\begin{array}{lcr}
(2,2,0): && 81.3 \pm 0.1\, \%\,, \\
(3,1,0): &&  9.3 \pm 0.1\, \%\,, \\
(1,3,0): &&  9.35 \pm 0.09\, \%\,, \\
(4,0,0): && 0.0006 \pm 0.0007\, \%\,, \\
(0,4,0): && 0.0004 \pm 0.0007\, \%\,. \\
\end{array}
\end{equation}
This represents the \emph{second main result} of the paper. Namely, in light of the discussion at the end of Subsection \ref{subsec:LengthScales}, if we have a QG model with a UV cutoff $l_\text{min}$ at the Planck scale, and if the model implements a path integral over both spacelike and timelike edge lengths, with the domain given by
\begin{equation} \label{eq:domainForSplitSig}
l_n \in ( l_p , \frac{10}{k} l_p] \cup (i l_p, i \frac{10}{k}l_p] \,,
\end{equation}
for any choice of $k\in(0,10)$, such a model generically \emph{predicts the probability of observing the split signature at any point in spacetime} to be roughly $80 \,\%$. The remaining $20\,\%$ are equally split between the Minkowski and negative Minkowski signature, while the Euclidean and negative Euclidean signatures have probability bounded from above by $0.0007\,\%$, i.e., almost zero.

If anything, this result suggests that one should \emph{not use} the mixed domain for edge lengths when constructing a model of QG, given that it would predict a very high probability of observing the split signature of the metric, which is obviously in contradiction with experiment. Also, in contrast to the spacelike/timelike case, this result is mostly independent on the choice of the cutoff $k$ --- across all values of the cutoff, the frequency of split signature varies roughly between $70\,\%$ and $80\,\%$, see Table \ref{tab3}.

The second feature of Figure \ref{fig:tab3} is a very pronounced symmetry between frequencies of Minkowski and negative Minkowski signatures, and between Euclidean and negative Euclidean signatures. This is of course expected. Namely, given the isomorphism between the spacelike and timelike domains discussed in the previous Subsection, in the case of the mixed domain such as (\ref{eq:domainForSplitSig}) the total dataset respects the symmetry $g_{\mu\nu} \to - g_{\mu\nu}$, which implies that one will observe equal frequencies for Minkowski and negative Minkowski signatures, and similarly for Euclidean and negative Euclidean. Obviously, the split signature itself is invariant with respect to $g_{\mu\nu} \to - g_{\mu\nu}$.

Finally, the third feature of Figure \ref{fig:tab3} is that both Euclidean signatures are severely suppressed, compared to the split signature and the two Minkowski signatures. This is not surprising, because the two Euclidean signatures require that
\begin{itemize}
\item[(i)] all $10$ edge lengths must be of the same type (either spacelike or timelike), and
\item[(ii)] the edge lengths must simultaneously satisfy all corresponding triangle, tetrahedral and $4$-simplex inequalities, expressed via the appropriate Cayley-Menger determinant inequalities (see our Theorem from Subsection \ref{subsec:CayleyMenger}).
\end{itemize}
Satisfying both requirements (i) and (ii), while possible, is rather rare if one samples the lengths from the domain (\ref{eq:domainForSplitSig}) using uniform distribution. In the case of the two Minkowski signatures, the requirement (i) is absent, while the corresponding requirement (ii) is less restrictive than for the Euclidean case, allowing the Minkowski and negative Minkowski signatures to reach frequencies of about $10\,\%$ each. In the case of the split signature, the requirement (i) is also absent, and as we can see from the data, the corresponding requirement (ii) is far less restrictive than for both Minkowski and Euclidean cases.

\subsection{Triangulations and topologies}

Let us now turn to the question how (if at all) the results obtained above depend on the choice of topology, and a particular triangulation of that topology. In order to address both questions using our numerical results, in addition to the results obtained for a set of disconnected $4$-simplices, we have also generated appropriate datasets for the topologies of the $4$-ball and the $4$-sphere, each represented using $10$ different triangulations. Tables \ref{tab:parametersBallTopology} and \ref{tab:parametersSphereTopology} list the parameters for each triangulation of each topology. While the number of tetrahedra is always fixed by the chosen number of $4$-simplices (and in the case of a $4$-ball, the chosen number of boundary tetrahedra), we can observe that each triangulation has a different number of vertices, edges and triangles. This reflects the fact that each triangulation has its own distinct combinatorial structure.

\begin{table}[H]
\begin{center}
\begin{tabular}{|c|c|c|c|c|c|c|} \hline
\ $4$-ball topology\ \ &\ boundary tetrahedra\ \ &\ vertices\ \ &\ edges\ \ &\ triangles\ \ &\ tetrahedra\ \ &\ $4$-simplices\ \ \\ \hline\hline
1 &\ 35\ 003\ \ &\ 26\ 102\ \ &\ 134\ 134\ \ &\ 275\ 536\ \ &\ 267\ 504\ \ &\ 100\ 001\ \ \\ \hline
2 &\ 35\ 003\ \ &\ 27\ 384\ \ &\ 137\ 980\ \ &\ 278\ 100\ \ &\ 267\ 504\ \ &\ 100\ 001\ \ \\ \hline
3 &\ 35\ 003\ \ &\ 27\ 523\ \ &\ 138\ 397\ \ &\ 278\ 378\ \ &\ 267\ 504\ \ &\ 100\ 001\ \ \\ \hline
4 &\ 35\ 003\ \ &\ 27\ 179\ \ &\ 137\ 365\ \ &\ 277\ 690\ \ &\ 267\ 504\ \ &\ 100\ 001\ \ \\ \hline
5 &\ 35\ 003\ \ &\ 25\ 197\ \ &\ 131\ 419\ \ &\ 273\ 726\ \ &\ 267\ 504\ \ &\ 100\ 001\ \ \\ \hline
6 &\ 35\ 003\ \ &\ 28\ 344\ \ &\ 140\ 860\ \ &\ 280\ 020\ \ &\ 267\ 504\ \ &\ 100\ 001\ \ \\ \hline
7 &\ 35\ 003\ \ &\ 26\ 473\ \ &\ 135\ 247\ \ &\ 276\ 278\ \ &\ 267\ 504\ \ &\ 100\ 001\ \ \\ \hline
8 &\ 35\ 003\ \ &\ 27\ 376\ \ &\ 137\ 956\ \ &\ 278\ 084\ \ &\ 267\ 504\ \ &\ 100\ 001\ \ \\ \hline
9 &\ 35\ 003\ \ &\ 28\ 391\ \ &\ 141\ 001\ \ &\ 280\ 114\ \ &\ 267\ 504\ \ &\ 100\ 001\ \ \\ \hline
10 &\ 35\ 003\ \ &\ 24\ 302\ \ &\ 128\ 734\ \ &\ 271\ 936\ \ &\ 267\ 504\ \ &\ 100\ 001\ \ \\ \hline
\end{tabular}
\end{center}
\caption{\label{tab:parametersBallTopology}Parameters of the $10$ triangulations corresponding to the $4$-ball topology.}
\end{table}

\begin{table}[H]
\begin{center}
\begin{tabular}{|c|c|c|c|c|c|} \hline
\ $4$-sphere topology\ \ &\ vertices\ \ &\ edges\ \ &\ triangles\ \ &\ tetrahedra\ \ &\ $4$-simplices\ \ \\ \hline\hline
1 &\ 15\ 823\ \ &\ 97\ 464\ \ &\ 231\ 646\ \ &\ 250\ 005\ \ &\ 100\ 002\ \ \\ \hline
2 &\ 19\ 088\ \ &\ 107\ 259\ \ &\ 238\ 176\ \ &\ 250\ 005\ \ &\ 100\ 002\ \ \\ \hline
3 &\ 16\ 393\ \ &\ 99\ 174\ \ &\ 232\ 786\ \ &\ 250\ 005\ \ &\ 100\ 002\ \ \\ \hline
4 &\ 15\ 652\ \ &\ 96\ 951\ \ &\ 231\ 304\ \ &\ 250\ 005\ \ &\ 100\ 002\ \ \\ \hline
5 &\ 19\ 372\ \ &\ 108\ 111\ \ &\ 238\ 744\ \ &\ 250\ 005\ \ &\ 100\ 002\ \ \\ \hline
6 &\ 18\ 121\ \ &\ 104\ 358\ \ &\ 236\ 242\ \ &\ 250\ 005\ \ &\ 100\ 002\ \ \\ \hline
7 &\ 16\ 177\ \ &\ 98\ 526\ \ &\ 232\ 354\ \ &\ 250\ 005\ \ &\ 100\ 002\ \ \\ \hline
8 &\ 15\ 995\ \ &\ 97\ 980\ \ &\ 231\ 990\ \ &\ 250\ 005\ \ &\ 100\ 002\ \ \\ \hline
9 &\ 15\ 177\ \ &\ 95\ 526\ \ &\ 230\ 354\ \ &\ 250\ 005\ \ &\ 100\ 002\ \ \\ \hline
10 &\ 18\ 926\ \ &\ 106\ 773\ \ &\ 237\ 852\ \ &\ 250\ 005\ \ &\ 100\ 002\ \ \\ \hline
\end{tabular}
\end{center}
\caption{\label{tab:parametersSphereTopology}Parameters of the $10$ triangulations corresponding to the $4$-sphere topology.}
\end{table}

For each of these $20$ triangulations, the numerical analysis analogous to the previous cases is performed --- for each of the three domains for edge lengths, after averaging over $10$ sample runs, the frequencies of all five signatures are expressed as percentages relative to a total number of $4$-simplices, along with the corresponding sample standard deviations $\sigma$, and the dependence of the frequencies on the choice of the cutoff $k\in\{0,\dots,9\}$ is then displayed on a graph.

\begin{figure}[H]
\begin{center}
\includegraphics[scale=0.5]{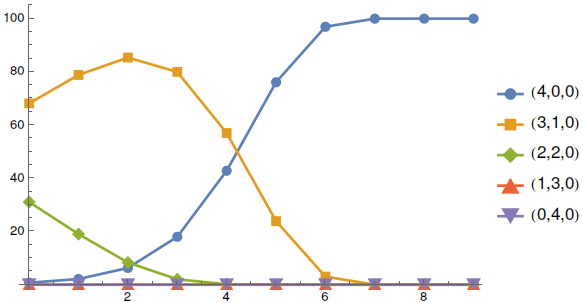}
\end{center}
\caption{\label{fig:tab4}Frequencies of various signatures, for the ball topology, triangulation number $3$ from Table \ref{tab:parametersBallTopology}, with spacelike domain for edge lengths. For each signature, and for each value of the cutoff $k$ on the horizontal axis, we plot the mean frequency as a percentage (\%). The corresponding sample standard deviations satisfy $\sigma \leq 0.4\,\%$ and are thus too small to be visible on the plot. The mean is obtained by averaging over $10$ sample runs, assigning a different set of edge lengths in each run. The data for this graph is taken from Table \ref{tab4}.}
\end{figure}

\begin{figure}[H]
\begin{center}
\includegraphics[scale=0.5]{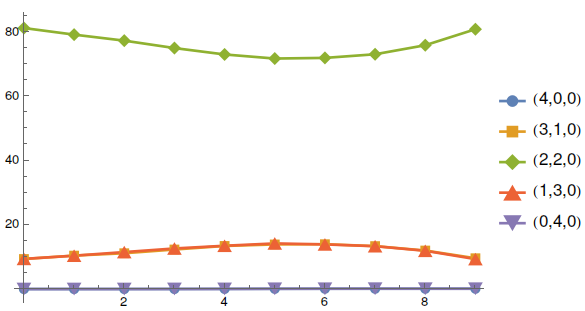}
\end{center}
\caption{\label{fig:tab5}Frequencies of various signatures, for the sphere topology, triangulation number $4$ from Table \ref{tab:parametersSphereTopology}, with mixed domain for edge lengths. For each signature, and for each value of the cutoff $k$ on the horizontal axis, we plot the mean frequency as a percentage (\%). The corresponding sample standard deviations satisfy $\sigma \leq 0.4\,\%$ and are thus too small to be visible on the plot. The mean is obtained by averaging over $10$ sample runs, assigning a different set of edge lengths in each run. The data for this graph is taken from Table \ref{tab5}.}
\end{figure}

In order to illustrate these results, in Figures \ref{fig:tab4} and \ref{fig:tab5} we plot the dependence of signature frequencies on the choice of the cutoff for two randomly chosen cases. The first corresponds to triangulation number $3$ of the $4$-ball topology with spacelike domain for edge lengths, while the second corresponds to the triangulation number $4$ of the $4$-sphere topology with the mixed domain for edge lengths.

Perhaps surprisingly, the results obtained for both topologies and all their triangulations are \emph{identical to those obtained for the disconnected topology}. The only difference (which is largely immaterial for our conclusions) is that the maximum value of the sample standard deviations is now $\sigma_\text{max} = 0.4\,\%$, as compared to $\sigma_\text{max} = 0.2\,\%$ obtained for the case of the disconnected topology. This can also be confirmed by comparing the numerical values of corresponding Tables \ref{tab4} and \ref{tab5} to the appropriate numerical values for the disconnected topology. The graphs obtained for all other choices of triangulations are the same as these two presented, and we omit them.

The fact that the results essentially do not depend on either the choice of a triangulation nor on the choice of the topology of the underlying spacetime manifold, represents our \emph{fourth main result}. Namely, the dependence of signature frequencies on the cutoff and the choice of domain for edge lengths appears to be a \emph{local property} of each $4$-simplex taken individually, regardless of the fact that this $4$-simplex may or may not be glued to other $4$-simplices in a triangulation.

This result is somewhat counter-intuitive, for the following reason. Namely, in a triangulation, a given $4$-simplex living in the bulk is connected to $5$ neighboring $4$-simplices, sharing one tetrahedron (and therefore $6$ edges) with each neighbor. Their common $6$ edges participate in the formula (\ref{eq:metricViaEdges}) for the metric tensors of both $4$-simplices, and consequently one would intuitively expect that this gives rise to correlations between the signatures of the two metrics. For example, if the common tetrahedron internally has Minkowski signature $(2,1,0)$, neither of the two $4$-simplices can have Euclidean signature. Nevertheless, our results demonstrate that any such correlations become invisible if one looks at a dataset consisting of a large number of $4$-simplices. The correlations either average out, or give rise to corrections in the signature frequencies that are below the standard deviations obtained from fluctuations of edge lengths themselves, and therefore remain invisible. 

At the end, let us mention that our study, strictly speaking, does not cover other possible choices of topologies of a $4$-dimensional manifold, such as a $4$-torus, or otherwise. While in principle any of these topologies should also be tested, the locality of the metric signature, as inferred from our data above, suggests that global properties of the manifold do not substantively contribute to signature frequencies, at least for the case of a large number of $4$-simplices in the triangulation.

\section{\label{secV}Conclusions}

In this work, we have investigated the \emph{ab initio} calculation of metric signatures in $4$-dimensional piecewise-flat models of quantum gravity, treating the signature of each simplex as a quantity determined by its edge lengths rather than imposing it a priori. We have established two complementary methods for evaluating the signature of a simplex --- direct computation of the eigenvalues of the metric tensor reconstructed from its edge lengths, and an algebraic method based on the signs of successive Cayley–Menger determinants. The agreement between these methods provides an independent consistency check for the numerical analysis. Using these techniques, we have performed a numerical investigation of the frequencies of the five possible nondegenerate metric signatures, considering different domains of edge lengths, ultraviolet cutoffs, triangulations, and topologies of the spacetime manifold.

Our results demonstrate that the statistical distribution of metric signatures depends strongly on the domain of integration chosen for the edge lengths. When all edges are restricted to be spacelike, the Minkowski signature can become strongly dominant. In particular, for a cutoff corresponding to edge lengths between the Planck scale and five times that scale, the frequency of the Minkowski signature reaches approximately $85\,\%$, with the remaining $15\,\%$ divided between Euclidean and split signature. Conversely, when both spacelike and timelike edges are allowed, the split signature becomes dominant, with probability between $72\,\%$ and $81\,\%$, for the sampled $4$-simplices across the whole range of the cutoff values. The purely timelike case yields the corresponding sign-reversed distribution isomorphic to the purely spacelike case. These results show that the choice of the domain of edge lengths plays a central role in determining the signature frequencies within the model, thus informing us about the preferred properties that a tentative QG model ought to have.

We have also found that the resulting signature frequencies are essentially unchanged across the disconnected, $4$-ball and $4$-sphere topologies examined in this work, and across a range of different triangulations considered for the latter two cases. This observation is consistent with the fact that the signature is determined locally by the edge lengths of each simplex and that the marginal distribution of those lengths is the same in all cases. Although neighboring simplices do share edges, and their signatures can therefore be correlated, such correlations do not appear to produce appreciable changes in the measured signature frequencies for the dataset studied in this work. Our results therefore suggest that the signature probabilities are largely insensitive to the global topology and combinatorial structure of the triangulation.

An important assumption of the analysis is that it is purely kinematical --- when evaluating the probabilities using our toy-model of QG (\ref{eq:ToyModelQG}), we have set the classical action to zero and sampled the edge lengths according to a prescribed uniform distribution. The reported frequencies therefore characterize the kinematical sector of the model, and should not be interpreted as universal probabilities for the physical spacetime signature or as predictions of a fully specified dynamical QG model. On the other hand, setting the classical action to zero renders the kinematical aspects of our model, and thus the obtained results, to be common baseline for a large class of QG models based on piecewise-flat spacetimes. The results demonstrate that a preference for a particular signature can emerge from the configuration space and measure alone, before any particular dynamics is taken into account. An important direction for future research is to consider the contribution of nontrivial dynamics, and investigate whether the resulting QG model preserves, suppresses, or enhances the obtained kinematical results for the signature. Further studies could also examine alternative measures and edge-length distributions, a broader range of triangulations and topologies, and the role of correlations between neighboring simplices. Such extensions would help test the robustness of the obtained predictions for the signature probabilities, and whether the spacetime signature can ultimately be understood as an emergent property of quantum geometry.

\acknowledgments


This research was supported by the Ministry of Science, Technological Development and Innovations of the Republic of Serbia (MNTRI).

\appendix

\section{\label{AppA}Signature evaluation from Cayley-Menger determinants}

Here we provide a proof of the Theorem stated in Subsection \ref{subsec:CayleyMenger}. To that end, let us first repeat the statement that should be proved.

\textbf{Theorem.}
\textit{Given a $D$-simplex with vertices denoted as $v_0,\dots,v_D$, introduce a sequence of $m$-simplices $\sigma_m$ constructed as subsimplices of the $D$-simplex, spanned over the following vertices:
\begin{equation}
\sigma_0 = \sspan(v_D)\,, \quad \sigma_1 = \sspan(v_{D-1},v_D)\,, \quad \dots \quad \sigma_m = \sspan(v_{D-m},\dots,v_D)\,, \quad \dots\quad \sigma_D = \sspan(v_0,\dots,v_D)\,.
\end{equation}
Assume that all corresponding Cayley-Menger determinants are nondegenerate, $C_m \neq 0$ for $m\in\{ 0,\dots,D \}$, and construct the sets
\begin{equation}
P = \Big\{  m\in\{1,\dots,D\}  \, \Big| \,  \frac{C_m}{C_{m-1}} < 0 \,  \Big\}\,, \qquad
N = \Big\{  m\in\{1,\dots,D\}  \, \Big| \,  \frac{C_m}{C_{m-1}} > 0 \,  \Big\}\,.
\end{equation}
Denote the numbers of elements of sets $P$ and $N$ as $|P|$ and $|N|$, respectively. Then, the signature of the metric corresponding to the $D$-simplex $\sigma_D$ is given by
\begin{equation} \label{eq:SignatureFromCMApp}
(p,n,z) = (|P|,|N|,0)\,.
\end{equation}
In other words, $p$ and $n$ are the \emph{number of alterations} and the \emph{number of preservations} of the signs of the following sequence Cayley-Menger determinants:
\begin{equation}
C_0 \equiv -1, \quad C_1\,, \quad C_2\,, \quad \dots\quad C_{D-1}\,,\qquad C_D\,.
\end{equation}
}

The proof relies on the relation (\ref{eq:relationCMandMetricDet}) between the Cayley-Menger determinant and the determinant of the metric, rewritten here in the form
\begin{equation} \label{eq:relationCMandMetricDetApp}
C_m = (-1)^{m+1} 2^m \Big( \prod_{j=D-m}^{D-1} | l_{Dj}^2 | \Big) \det g_{\mu\nu}^{(m)}\,, \qquad m\in\{ 1,\dots,D \}\,.
\end{equation}
Here we have added a superscript $m$ on the metric $g_{\mu\nu}^{(m)}$, in order to keep track of the fact that this is a metric corresponding to the $m$-simplex $\sigma_m$. Start by choosing any $m \geq 2$, and rewrite the metric tensor in the following block-matrix form:
\begin{equation} \label{eq:MetricBlockForm}
\left[g_{\alpha\beta}^{(m)}\right] = \left[
\begin{array}{c|ccc}
 c^{(m)}  & & b_\nu^{(m)} &  \\ \hline
 & & & \\
b_\mu^{(m)} & & g_{\mu\nu}^{(m-1)} &  \\ 
 & & & \\
\end{array}
\right]\,, \qquad
\begin{array}{l}
  \alpha,\beta \in \{D-m,\dots,D-1 \}\,, \\
  \mu,\nu \in \{D-m+1,\dots,D-1 \}\,. \\
\end{array}
\end{equation}
The first step is to establish that the block-matrix $g_{\mu\nu}^{(m-1)}$, sitting inside the metric $g_{\alpha\beta}^{(m)}$, is indeed the metric of the simplex $\sigma_{m-1}$. To see this, recall that according to (\ref{eq:metricViaEdgesDdim}), for the $m$-simplex $\sigma_m = \sspan(v_{D-m},\dots,v_D)$ we have
\begin{equation}
g_{\alpha\beta}^{(m)} = \frac{l_{D\alpha}^2 + l_{D\beta}^2 - l_{\alpha\beta}^2}{2 |l_{D\alpha}| |l_{D\beta}|}\,, \qquad \alpha,\beta\in \{D-m,\dots,D-1\} \,.
\end{equation}
On the other hand, the $(m-1)$-simplex $\sigma_{m-1} = \sspan(v_{D-m+1},\dots,v_D)$ is obtained from $\sigma_m$ by removing the vertex $v_{D-m}$, so for $\sigma_{m-1}$ we write the analogous formula for the metric,
\begin{equation}
g_{\mu\nu}^{(m-1)} = \frac{l_{D\mu}^2 + l_{D\nu}^2 - l_{\mu\nu}^2}{2 |l_{D\mu}| |l_{D\nu}|}\,, \qquad \mu,\nu\in \{D-m+1,\dots,D-1\} \,,
\end{equation}
which is precisely the expression for the block in (\ref{eq:MetricBlockForm}) obtained by removing the first row and the first column, since these correspond to the vertex $v_{D-m}$.

The second step is to construct the $1$-dimensional Schur complement of the metric (\ref{eq:MetricBlockForm}). Since the metric $g_{\mu\nu}^{(m-1)}$ is by assumption nondegenerate and thus invertible, the Schur complement can be written in the form
\begin{equation}
d^{(m)} =  c^{(m)} - b_\mu^{(m)} g^{\mu\nu}_{(m-1)} b_\nu^{(m)}\,.
\end{equation}
Introducing the lower triangular matrix $P_{(m)}$ and its transpose $P_{(m)}^T$ as
\begin{equation}
\left[ P_{(m)}{}^\beta{}_\gamma \right] = \left[
\begin{array}{c|ccc}
  1 & \;\;\; & 0 & \;\;\;  \\ \hline
  & & &  \\
-g^{\nu\lambda}_{(m-1)} b_\lambda^{(m)}  & & \delta^\nu_\rho &  \\
  & & &  \\
\end{array}
\right]\,, \qquad
\left[ P^T_{(m)\delta}{}^\alpha \right] = \left[
\begin{array}{c|ccc}
  \;1\; & \;\; & -g^{\mu\tau}_{(m-1)} b_\tau^{(m)} & \;\;  \\ \hline
  & & &  \\
0  &  & \delta^\mu_\sigma &  \\
  & & &  \\
\end{array}
\right]\,, 
\end{equation}
where we have separated the indices as
\begin{equation}
\alpha \to (1,\mu)\,, \qquad \beta \to (1,\nu)\,, \qquad \gamma\to (1,\rho)\,, \qquad \delta\to (1,\sigma)\,,
\end{equation}
the main property of the Schur complement gives us the following decomposition:
\begin{equation} \label{eq:SchurDecomposition}
P^T_{(m)} g^{(m)} P_{(m)}  = \left[ P^T_{(m)\delta}{}^\alpha g_{\alpha\beta}^{(m)} P_{(m)}{}^\beta{}_\gamma \right] = \left[
\begin{array}{c|ccc}
 d^{(m)} & & 0 & \\ \hline
 & & & \\
0 & & g_{\sigma\rho}^{(m-1)} & \\
 & & & \\
\end{array}
\right] = \left[ d^{(m)} \right] \oplus \left[ g_{\sigma\rho}^{(m-1)} \right] = d^{(m)} \oplus g^{(m-1)}\,.
\end{equation}
The most important result one should deduce from this decomposition is that according to the Sylvester's law of inertia, the left-hand side has precisely the signature of $g_{\alpha\beta}^{(m)}$, denoted $(p_m,n_m,0)$, while the right-hand side has the signature of $g_{\sigma\rho}^{(m-1)}$, denoted $(p_{m-1},n_{m-1},0)$ plus the signature of the $1\times 1$ matrix $d^{(m)}$, which can be either $(1,0,0)$ or $(0,1,0)$, depending on whether the Schur complement $d^{(m)}$ is positive or negative, respectively (by assumption of nondegeneracy, the Schur complement can never be zero). In other words, we have
\begin{equation} \label{eq:signatureRecursionFormula}
(p_m,n_m,0) = \left\{
\begin{array}{lclcl}
(1+p_{m-1}, n_{m-1},0) & & \text{ if } & &  d^{(m)} > 0\,, \\
(p_{m-1}, 1+ n_{m-1},0) & & \text{ if } & & d^{(m)} < 0\,. \\
\end{array}
\right.
\end{equation}

The third step is to determine when is the Schur complement positive, and when it is negative. This is done by taking the determinant of the decomposition (\ref{eq:SchurDecomposition}). Given that $P_{(m)}$ and $P_{(m)}^T$ are triangular matrices with unit diagonals, we have that $\det P_{(m)} = \det P_{(m)}^T = 1$, so we obtain
\begin{equation}
\det g_{\alpha\beta}^{(m)} = d^{(m)} \det g_{\sigma\rho}^{(m-1)}\,,
\end{equation}
where it is now obvious that $d^{(m)} \neq 0$ since both metric tensors are nondegenerate. Solving for the Schur complement, and applying (\ref{eq:relationCMandMetricDetApp}), we obtain:
\begin{equation}
d^{(m)} = \frac{\det g_{\alpha\beta}^{(m)}}{\det g_{\sigma\rho}^{(m-1)}} =  - \frac{1}{2} \; \frac{1}{| l^2_{D,D-m} | }\;  \frac{C_m}{C_{m-1}} \,.
\end{equation}
We conclude that the Schur complement $d^{(m)}$ is positive if the ratio of Cayley-Menger determinants $C_m / C_{m-1}$ is negative, and vice versa. Substituting this into (\ref{eq:signatureRecursionFormula}), we obtain the recursion formula for the metric signature in the form:
\begin{equation} \label{eq:signatureRecursionFormulaFinal}
  (p_m,n_m,0) = \left\{
\begin{array}{lclcl}
(1+p_{m-1}, n_{m-1},0) & & \text{ if } & & \ds  \frac{C_m}{C_{m-1}} < 0\,, \vphantom{\ds\int_A} \\
(p_{m-1}, 1+ n_{m-1},0) & & \text{ if } & & \ds \frac{C_m}{C_{m-1}} > 0\,. \vphantom{\ds\int^A} \\
\end{array}
\right.
\end{equation}
We see that the number of positive eigenvalues increases by one whenever the Cayley-Menger determinant of the simplex $\sigma_m$ flips its sign relative to the corresponding subsimplex $\sigma_{m-1}$, while the number of negative eigenvalues increases by one whenever the Cayley-Menger determinant preserves its sign. Looking at the whole sequence of simplices,
\begin{equation}
\sigma_0 = \sspan(v_D)\,, \quad \sigma_1 = \sspan(v_{D-1},v_D)\,, \quad \dots \quad \sigma_m = \sspan(v_{D-m},\dots,v_D)\,, \quad \dots\quad \sigma_D = \sspan(v_0,\dots,v_D)\,,
\end{equation}
we can evaluate a corresponding sequence of Cayley-Menger determinants,
\begin{equation}
C_0 \equiv -1, \quad C_1\,, \quad C_2\,, \quad \dots\quad C_{D-1}\,,\qquad C_D\,,
\end{equation}
and introduce the sets $P$ and $N$ that count the number of sign alterations and sign preservations, respectively:
\begin{equation}
P = \Big\{  m\in\{1,\dots,D\}  \, \Big| \,  \frac{C_m}{C_{m-1}} < 0 \,  \Big\}\,, \qquad
N = \Big\{  m\in\{1,\dots,D\}  \, \Big| \,  \frac{C_m}{C_{m-1}} > 0 \,  \Big\}\,.
\end{equation}
The total number of sign flips corresponds to the total number of elements $|P|$ in $P$, while the total number of sign preservations corresponds to the total number of elements $|N|$ in $N$. Since the $0$-simplex $\sigma_0$ is just a single vertex, its signature is by definition given as $(p_0,n_0,z_0) = (0,0,0)$. Then, as we move through the sequence of simplices up to $\sigma_D$, we observe from the recursion formula (\ref{eq:signatureRecursionFormulaFinal}) that after $|P|$ sign alterations and $|N|$ sign preservations, we obtain that the metric signature corresponding to the $D$-simplex $\sigma_D$ is given as
\begin{equation}
(p_D,n_D,z_D) = (|P|,|N|,0)\,,
\end{equation}
which is identical to (\ref{eq:SignatureFromCMApp}). This completes the proof.

\section{\label{AppB}Tables with numerical data}

Here we provide the numerical data corresponding to the graphs displayed in Figures \ref{fig:tab1}, \ref{fig:tab2}, \ref{fig:tab3}, \ref{fig:tab4} and \ref{fig:tab5}, from the main text.

\begin{table}[H]
\begin{center}
\begin{tabular}{|cc|r@{.}l|r@{.}l|r@{.}l|r@{.}l|r@{.}l|r@{.}l|r@{.}l|r@{.}l|c|c|} \hline
\ \ \ \ cutoff: &  & \multicolumn{2}{c|}{$k=0$} &\multicolumn{2}{c|}{$k=1$} & \multicolumn{2}{c|}{$k=2$} & \multicolumn{2}{c|}{$k=3$} & \multicolumn{2}{c|}{$k=4$} & \multicolumn{2}{c|}{$k=5$} & \multicolumn{2}{c|}{$k=6$} & \multicolumn{2}{c|}{$k=7$} & $k=8$ & $k=9$ \\ \hline \hline
\multirow{2}{*}{$(4,0,0)$} & \% & 0 & 75 & 2 & 11\ \  & 6 & 36 & 18 & 0 & 42 & 7 & 76 & 0 &\ 97 & 01 &\ 99 & 996\ \ &\ 100 \ \  &\ 100 \ \  \\
                                         &$\sigma$ & 0 & 02 & 0 & 03 & 0 & 06 & 0 & 1 & 0 & 1 & 0 & 2 & 0 & 06 & 0 & 001 & 0 & 0  \\ \hline
\multirow{2}{*}{$(3,1,0)$} &\% &\ 67 & 99 &\ 79 & 0 &\ 85 & 34 &\ 79 & 9 &\ 57 & 1 &\ 24 & 0 & 2 & 99\ \  & 0 & 004 & 0 & 0  \\
                                        &$\sigma$ &0 & 09 & 0 & 1 & 0 & 07 & 0 & 1 & 0 & 1 & 0 & 2 & 0 & 06 & 0 & 001 & 0 & 0  \\\hline
\multirow{2}{*}{$(2,2,0)$} &\% &31 & 26 & 18 & 8 & 8 & 30\ \ & 2 & 08 & 0 & 16\ \  & 0 & 002\ \  & \multicolumn{2}{c|}{0} & \multicolumn{2}{c|}{0} & 0 & 0  \\
                                        &$\sigma$ &0 & 08 & 0 & 1 & 0 & 06 & 0 & 07\ \  & 0 & 01 & 0 & 001 & \multicolumn{2}{c|}{0} & \multicolumn{2}{c|}{0} & 0 & 0  \\ \hline
\multirow{2}{*}{$(1,3,0)$} &\% &0 & 003\ \ & \multicolumn{2}{c|}{0} & \multicolumn{2}{c|}{0} & \multicolumn{2}{c|}{0} & \multicolumn{2}{c|}{0} & \multicolumn{2}{c|}{0} & \multicolumn{2}{c|}{0} & \multicolumn{2}{c|}{0} & 0 & 0  \\
                                        &$\sigma$ &0 & 002 & \multicolumn{2}{c|}{0} & \multicolumn{2}{c|}{0} & \multicolumn{2}{c|}{0} & \multicolumn{2}{c|}{0} & \multicolumn{2}{c|}{0} & \multicolumn{2}{c|}{0} & \multicolumn{2}{c|}{0} & 0 & 0  \\\hline
\multirow{2}{*}{$(0,4,0)$} &\% &\multicolumn{2}{c|}{0} & \multicolumn{2}{c|}{0} & \multicolumn{2}{c|}{0} & \multicolumn{2}{c|}{0} & \multicolumn{2}{c|}{0} & \multicolumn{2}{c|}{0} & \multicolumn{2}{c|}{0} & \multicolumn{2}{c|}{0} & 0 & 0  \\
                                         &$\sigma$ &\multicolumn{2}{c|}{0} & \multicolumn{2}{c|}{0} & \multicolumn{2}{c|}{0} & \multicolumn{2}{c|}{0} & \multicolumn{2}{c|}{0} & \multicolumn{2}{c|}{0} & \multicolumn{2}{c|}{0} & \multicolumn{2}{c|}{0} & 0 & 0  \\ \hline
\end{tabular}
\end{center}
\caption{\label{tab1}Frequencies of various signatures, for the disconnected topology with spacelike domain for edge lengths. For each signature, and for each value of the cutoff $k$, we quote the mean frequency as a percentage (\%), as well as the sample standard deviation around that mean ($\sigma$). The mean and the standard deviation are obtained by averaging over $10$ sample runs, assigning a different set of edge lengths in each run. The data from this table is depicted in Figure \ref{fig:tab1}.}
\end{table}

\begin{table}[H]
\begin{center}
\begin{tabular}{|cc|r@{.}l|r@{.}l|r@{.}l|r@{.}l|r@{.}l|r@{.}l|r@{.}l|r@{.}l|c|c|} \hline
\ \ \ \ cutoff: &  & \multicolumn{2}{c|}{$k=0$} &\multicolumn{2}{c|}{$k=1$} & \multicolumn{2}{c|}{$k=2$} & \multicolumn{2}{c|}{$k=3$} & \multicolumn{2}{c|}{$k=4$} & \multicolumn{2}{c|}{$k=5$} & \multicolumn{2}{c|}{$k=6$} & \multicolumn{2}{c|}{$k=7$} &\ $k=8$\ \  &\ $k=9$\ \  \\ \hline \hline
\multirow{2}{*}{$(4,0,0)$} & \% & \multicolumn{2}{c|}{0} & \multicolumn{2}{c|}{0} & \multicolumn{2}{c|}{0} & \multicolumn{2}{c|}{0} & \multicolumn{2}{c|}{0} & \multicolumn{2}{c|}{0} & \multicolumn{2}{c|}{0} & \multicolumn{2}{c|}{0} & 0 & 0 \\
                                         &$\sigma$ & \multicolumn{2}{c|}{0} & \multicolumn{2}{c|}{0} & \multicolumn{2}{c|}{0} & \multicolumn{2}{c|}{0} & \multicolumn{2}{c|}{0} & \multicolumn{2}{c|}{0} & \multicolumn{2}{c|}{0} & \multicolumn{2}{c|}{0} & 0 & 0 \\ \hline
\multirow{2}{*}{$(3,1,0)$} & \% & 0 & 002\ \  & \multicolumn{2}{c|}{0} & \multicolumn{2}{c|}{0} & \multicolumn{2}{c|}{0} & \multicolumn{2}{c|}{0} & \multicolumn{2}{c|}{0} & \multicolumn{2}{c|}{0} & \multicolumn{2}{c|}{0} & 0 & 0 \\
                                         &$\sigma$ & 0 & 002 & \multicolumn{2}{c|}{0} & \multicolumn{2}{c|}{0} & \multicolumn{2}{c|}{0} & \multicolumn{2}{c|}{0} & \multicolumn{2}{c|}{0} & \multicolumn{2}{c|}{0} & \multicolumn{2}{c|}{0} & 0 & 0 \\ \hline
\multirow{2}{*}{$(2,2,0)$} & \% & 31 & 2 & 18 & 9 & 8 & 33 & 2 & 07\ \  & 0 & 16\ \  & 0 & 001\ \  & \multicolumn{2}{c|}{0} & \multicolumn{2}{c|}{0} & 0 & 0 \\
                                         &$\sigma$ & 0 & 1 & 0 & 1 & 0 & 08 & 0 & 04 & 0 & 02 & 0 & 001 & \multicolumn{2}{c|}{0} & \multicolumn{2}{c|}{0} & 0 & 0 \\ \hline
\multirow{2}{*}{$(1,3,0)$} & \% &\ 68 & 1 &\ 79 & 0 &\ 85 & 38 &\ 79 & 9 &\ 57 & 2 &\ 23 & 9 & 3 & 00 & 0 & 003 & 0 & 0 \\
                                         &$\sigma$ & 0 & 1 & 0 & 1 & 0 & 08 & 0 & 2 & 0 & 2 & 0 & 1 & 0 & 05 & 0 & 001 & 0 & 0 \\ \hline
\multirow{2}{*}{$(0,4,0)$} & \% & 0 & 74 & 2 & 12 & 6 & 29 & 18 & 0 & 42 & 6 &\ 76 & 1 &\ 97 & 00 &\ 99 & 997\ \  & 100 & 100 \\
                                         &$\sigma$ & 0 & 02 & 0 & 04\ \  & 0 & 08\ \  & 0 & 2 & 0 & 2 & 0 & 1 & 0 & 05\ \  & 0 & 001 & 0 & 0 \\ \hline
\end{tabular}
\end{center}
\caption{\label{tab2}Frequencies of various signatures, for the disconnected topology with timelike domain for edge lengths. For each signature, and for each value of the cutoff $k$, we quote the mean frequency as a percentage (\%), as well as the sample standard deviation around that mean ($\sigma$). The mean and the standard deviation are obtained by averaging over $10$ sample runs, assigning a different set of edge lengths in each run. The data from this table is depicted in Figure \ref{fig:tab2}.}
\end{table}

\begin{table}[H]
\begin{center}
\begin{tabular}{|cc|r@{.}l|r@{.}l|r@{.}l|r@{.}l|r@{.}l|r@{.}l|r@{.}l|r@{.}l|r@{.}l|r@{.}l|} \hline
\ \ \ \ cutoff: &  & \multicolumn{2}{c|}{$k=0$} &\multicolumn{2}{c|}{$k=1$} & \multicolumn{2}{c|}{$k=2$} & \multicolumn{2}{c|}{$k=3$} & \multicolumn{2}{c|}{$k=4$} & \multicolumn{2}{c|}{$k=5$} & \multicolumn{2}{c|}{$k=6$} & \multicolumn{2}{c|}{$k=7$} & \multicolumn{2}{c|}{$k=8$}  & \multicolumn{2}{c|}{$k=9$}  \\ \hline \hline
\multirow{2}{*}{$(4,0,0)$} & \% & 0 & 0006\ \  & 0 & 001\ \  & 0 & 007\ \  & 0 & 018\ \  & 0 & 042\ \  & 0 & 075\ \  & 0 & 094\ \  & 0 & 090\ \  & 0 & 10\ \  & 0 & 10 \\
                                         &$\sigma$ & 0 & 0007 & 0 & 001 & 0 & 003 & 0 & 003 & 0 & 004 & 0 & 007 & 0 & 009 & 0 & 007 & 0 & 01 & 0 & 02 \\ \hline
\multirow{2}{*}{$(3,1,0)$} & \% & 9 & 3 & 10 & 4 & 11 & 4 & 12 & 48 & 13 & 4 & 14 & 0 & 13 & 91 & 13 & 3 & 11 & 97 & 9 & 5 \\
                                         &$\sigma$ & 0 & 1 & 0 & 1 & 0 & 1 & 0 & 08 & 0 & 1 & 0 & 1 & 0 & 09 & 0 & 1 & 0 & 05 & 0 & 1 \\ \hline
\multirow{2}{*}{$(2,2,0)$} & \% &\ 81 & 3 &\ 79 & 3 &\ 77 & 1 &\ 75 & 0 &\ 73 & 02 &\ 71 & 9 &\ 72 & 05 &\ 73 & 2 &\ 75 & 9 &\ 80 & 9 \\
                                         &$\sigma$ & 0 & 1 & 0 & 2 & 0 & 2 & 0 & 1 & 0 & 2 & 0 & 1 & 0 & 07 & 0 & 1 & 0 & 1 & 0 & 1 \\ \hline
\multirow{2}{*}{$(1,3,0)$} & \% & 9 & 35 & 10 & 3 & 11 & 5 & 12 & 5 & 13 & 46 & 13 & 9 & 13 & 84 & 13 & 3 & 11 & 9 & 9 & 5 \\
                                         &$\sigma$ & 0 & 09 & 0 & 1 & 0 & 1 & 0 & 1 & 0 & 07 & 0 & 1 & 0 & 08 & 0 & 1 & 0 & 1 & 0 & 1 \\ \hline
\multirow{2}{*}{$(0,4,0)$} & \% & 0 & 0004 & 0 & 001 & 0 & 007 & 0 & 017 & 0 & 04 & 0 & 074 & 0 & 099 & 0 & 092 & 0 & 10 & 0 & 096\ \  \\
                                         &$\sigma$ & 0 & 0007 & 0 & 001 & 0 & 003 & 0 & 002 & 0 & 01 & 0 & 007 & 0 & 009 & 0 & 008 & 0 & 01 & 0 & 005\ \  \\ \hline
\end{tabular}
\end{center}
\caption{\label{tab3}Frequencies of various signatures, for the disconnected topology with mixed domain for edge lengths. For each signature, and for each value of the cutoff $k$, we quote the mean frequency as a percentage (\%), as well as the sample standard deviation around that mean ($\sigma$). The mean and the standard deviation are obtained by averaging over $10$ sample runs, assigning a different set of edge lengths in each run. The data from this table is depicted in Figure \ref{fig:tab3}.}
\end{table}

\begin{table}[H]
\begin{center}
\begin{tabular}{|cc|r@{.}l|r@{.}l|r@{.}l|r@{.}l|r@{.}l|r@{.}l|r@{.}l|r@{.}l|r@{.}l|r@{.}l|} \hline
\ \ \ \ cutoff: &  & \multicolumn{2}{c|}{$k=0$} &\multicolumn{2}{c|}{$k=1$} & \multicolumn{2}{c|}{\ $k=2$\ \ } & \multicolumn{2}{c|}{$k=3$} & \multicolumn{2}{c|}{$k=4$} & \multicolumn{2}{c|}{$k=5$} & \multicolumn{2}{c|}{$k=6$} & \multicolumn{2}{c|}{$k=7$} & \multicolumn{2}{c|}{\ $k=8$\ \ }  & \multicolumn{2}{c|}{\ $k=9$\ \ }  \\ \hline \hline
\multirow{2}{*}{$(4,0,0)$} & \% & 0 & 75 & 2 & 15\ \ & 6 & 3 & 18 & 0 & 42 & 8 & 76 & 1 &\ 96 & 95\ \ &\ 99 & 999\ \ & \multicolumn{2}{c|}{100} & \multicolumn{2}{c|}{100} \\
                                         &$\sigma$ & 0 & 05 & 0 & 07 & 0 & 2 & 0 & 3 & 0 & 4 & 0 & 3 & 0 & 08 & 0 & 002 & \multicolumn{2}{c|}{0} & \multicolumn{2}{c|}{0} \\ \hline
\multirow{2}{*}{$(3,1,0)$} & \% &\ 68 & 2 &\ 78 & 8 &\ 85 & 4 &\ 80 & 0 &\ 57 & 0 &\ 23 & 9 & 3 & 05 & 0 & 003 & \multicolumn{2}{c|}{0} & \multicolumn{2}{c|}{0} \\
                                         &$\sigma$ & 0 & 2 & 0 & 2 & 0 & 1 & 0 & 3 & 0 & 4 & 0 & 3 & 0 & 08 & 0 & 002 & \multicolumn{2}{c|}{0} & \multicolumn{2}{c|}{0} \\ \hline
\multirow{2}{*}{$(2,2,0)$} & \% & 31 & 1 & 19 & 0 & 8 & 3 & 2 & 04\ \ & 0 & 16\ \ & 0 & 0012\ \ & \multicolumn{2}{c|}{0} & \multicolumn{2}{c|}{0} & \multicolumn{2}{c|}{0} & \multicolumn{2}{c|}{0} \\
                                         &$\sigma$ & 0 & 2 & 0 & 3 & 0 & 2 & 0 & 03 & 0 & 01 & 0 & 0008 & \multicolumn{2}{c|}{0} & \multicolumn{2}{c|}{0} & \multicolumn{2}{c|}{0} & \multicolumn{2}{c|}{0} \\ \hline
\multirow{2}{*}{$(1,3,0)$} & \% & 0 & 003\ \  & \multicolumn{2}{c|}{0} & \multicolumn{2}{c|}{0} & \multicolumn{2}{c|}{0} & \multicolumn{2}{c|}{0} & \multicolumn{2}{c|}{0} & \multicolumn{2}{c|}{0} & \multicolumn{2}{c|}{0} & \multicolumn{2}{c|}{0} & \multicolumn{2}{c|}{0} \\
                                         &$\sigma$ & 0 & 002 & \multicolumn{2}{c|}{0} & \multicolumn{2}{c|}{0} & \multicolumn{2}{c|}{0} & \multicolumn{2}{c|}{0} & \multicolumn{2}{c|}{0} & \multicolumn{2}{c|}{0} & \multicolumn{2}{c|}{0} & \multicolumn{2}{c|}{0} & \multicolumn{2}{c|}{0} \\ \hline
\multirow{2}{*}{$(0,4,0)$} & \% & \multicolumn{2}{c|}{0} & \multicolumn{2}{c|}{0} & \multicolumn{2}{c|}{0} & \multicolumn{2}{c|}{0} & \multicolumn{2}{c|}{0} & \multicolumn{2}{c|}{0} & \multicolumn{2}{c|}{0} & \multicolumn{2}{c|}{0} & \multicolumn{2}{c|}{0} & \multicolumn{2}{c|}{0} \\
                                         &$\sigma$ & \multicolumn{2}{c|}{0} & \multicolumn{2}{c|}{0} & \multicolumn{2}{c|}{0} & \multicolumn{2}{c|}{0} & \multicolumn{2}{c|}{0} & \multicolumn{2}{c|}{0} & \multicolumn{2}{c|}{0} & \multicolumn{2}{c|}{0} & \multicolumn{2}{c|}{0} & \multicolumn{2}{c|}{0} \\ \hline
\end{tabular}
\end{center}
\caption{\label{tab4}Frequencies of various signatures, for the ball topology, triangulation number $3$ from Table \ref{tab:parametersBallTopology}, with spacelike domain for edge lengths. For each signature, and for each value of the cutoff $k$, we quote the mean frequency as a percentage (\%), as well as the sample standard deviation around that mean ($\sigma$). The mean and the standard deviation are obtained by averaging over $10$ sample runs, assigning a different set of edge lengths in each run. The data from this table is depicted in Figure \ref{fig:tab4}.}
\end{table}

\begin{table}[H]
\begin{center}
\begin{tabular}{|cc|r@{.}l|r@{.}l|r@{.}l|r@{.}l|r@{.}l|r@{.}l|r@{.}l|r@{.}l|r@{.}l|r@{.}l|} \hline
\ \ \ \ cutoff: &  & \multicolumn{2}{c|}{\ $k=0$\ \ } &\multicolumn{2}{c|}{\ $k=1$\ \ } & \multicolumn{2}{c|}{\ $k=2$\ \ } & \multicolumn{2}{c|}{\ $k=3$\ \ } & \multicolumn{2}{c|}{\ $k=4$\ \ } & \multicolumn{2}{c|}{$k=5$} & \multicolumn{2}{c|}{\ $k=6$\ \ } & \multicolumn{2}{c|}{\ $k=7$\ \ } & \multicolumn{2}{c|}{\ $k=8$\ \ }  & \multicolumn{2}{c|}{\ $k=9$\ \ }  \\ \hline \hline
\multirow{2}{*}{$(4,0,0)$} & \% & 0 & 0007\ \ & 0 & 002 & 0 & 006 & 0 & 017 & 0 & 04 & 0 & 07 & 0 & 10 & 0 & 10 & 0 & 10 & 0 & 11 \\
                                         &$\sigma$ & 0 & 0008 & 0 & 002 & 0 & 003 & 0 & 003 & 0 & 01 & 0 & 01 & 0 & 02 & 0 & 01 & 0 & 01 & 0 & 01 \\ \hline
\multirow{2}{*}{$(3,1,0)$} & \% & 9 & 4 & 10 & 4 & 11 & 1 & 12 & 3 & 13 & 4 & 13 & 9 & 14 & 0 & 13 & 4 & 12 & 0 & 9 & 6 \\
                                         &$\sigma$ & 0 & 2 & 0 & 4 & 0 & 3 & 0 & 3 & 0 & 4 & 0 & 4 & 0 & 3 & 0 & 4 & 0 & 4 & 0 & 3 \\ \hline
\multirow{2}{*}{$(2,2,0)$} & \% &\ 81 & 3 &\ 79 & 2 &\ 77 & 3 &\ 75 & 0 &\ 73 & 0 &\ 71 & 8 &\ 72 & 0 &\ 73 & 1 &\ 75 & 9 &\ 80 & 9 \\
                                         &$\sigma$ & 0 & 2 & 0 & 1 & 0 & 1 & 0 & 2 & 0 & 2 & 0 & 2 & 0 & 2 & 0 & 2 & 0 & 2 & 0 & 2 \\ \hline
\multirow{2}{*}{$(1,3,0)$} & \% & 9 & 4 & 10 & 4 & 11 & 5 & 12 & 6 & 13 & 5 & 14 & 2 & 13 & 9 & 13 & 3 & 11 & 9 & 9 & 3 \\
                                         &$\sigma$ & 0 & 1 & 0 & 3 & 0 & 3 & 0 & 2 & 0 & 3 & 0 & 4 & 0 & 3 & 0 & 3 & 0 & 3 & 0 & 2 \\ \hline
\multirow{2}{*}{$(0,4,0)$} & \% & 0 & 0013 & 0 & 002\ \ & 0 & 005\ \ & 0 & 021\ \ & 0 & 04\ \ & 0 & 08\ \ & 0 & 09\ \ & 0 & 11\ \ & 0 & 10\ \ & 0 & 10\ \ \\
                                         &$\sigma$ & 0 & 0008 & 0 & 002 & 0 & 002 & 0 & 005 & 0 & 01 & 0 & 01 & 0 & 01 & 0 & 02 & 0 & 01 & 0 & 01 \\ \hline
\end{tabular}
\end{center}
\caption{\label{tab5}Frequencies of various signatures, for the sphere topology, triangulation number $4$ from Table \ref{tab:parametersSphereTopology}, with mixed domain for edge lengths. For each signature, and for each value of the cutoff $k$, we quote the mean frequency as a percentage (\%), as well as the sample standard deviation around that mean ($\sigma$). The mean and the standard deviation are obtained by averaging over $10$ sample runs, assigning a different set of edge lengths in each run. The data from this table is depicted in Figure \ref{fig:tab5}.}
\end{table}

\end{document}